\documentstyle[12pt,a4wide]{article}
\normalsize
\def\simless{\mathbin{\lower 3pt\hbox{$\rlap{\raise 5pt\hbox{$\char'074$}}
\mathchar"7218$}}}
\def\simgreat{\mathbin{\lower 3pt\hbox{$\rlap{\raise 5pt \hbox{$\char'076$}}
\mathchar"7218$}}}
\catcode`@=11

\def\beqra{\begin{eqnarray}} \def\eeqra{\end{eqnarray}}
\def\beq{\begin{equation}}      \def\eeq{\end{equation}}

\def\fo{\hbox{{1}\kern-.25em\hbox{l}}}

\def\ch{\@startsection{section}{1}{\z@}{-3ex plus-1ex minus-.2ex}%
        {2ex plus.2ex}{\large\sc}}

\def\; \lapp \;{\raisebox{-.4ex}{\rlap{$\sim$}} \raisebox{.4ex}{$<$}}

\def\con{\ifmmode \hbox{\bf*} \else{\bf*}\fi}   % conjugation
\def\scon{\ifmmode \hbox{\footnotesize\rm\bf*} \else{\footnotesize\rm\bf*}\fi}

\def\0#1{\relax\ifmmode\mathaccent"7017{#1}%    % puts a little circle atop,
        \else\accent23#1\relax\fi}              % as a halo of a saint

\def\eslash{\not{\hbox{\kern-2pt $E$}}}

\catcode`@=12

\begin{document}
\begin{titlepage}
\hoffset=0.4cm
\voffset=-1truecm
\normalsize
\def\ni{{\bar {N_i}}}    \def\nj{{\bar {N_j}}}   \def\n3{{\bar {N_3}}}
\def\li{\lambda_i}    \def\lj{\lambda_j}   \def\l3{\lambda_3}
\def\hn{h^\nu}       \def\hnij{h^\nu_{ij}}
\baselineskip=5pt
\begin{flushright}
DFPD 93/TH/06
\end{flushright}
\begin{flushright}
UTS-DFT-93-1
\end{flushright}
\vspace{24pt}
\begin{center}

{\Large {\bf  COVARIANT ANOMALIES AND FUNCTIONAL DETERMINANTS}}
\end{center}
\vspace{24pt}
\centerline{\large Luca Griguolo}
\vskip 0.2 cm
\centerline{\it SISSA, via Beirut 2-34100 Trieste, Italy}
\centerline{\it INFN, Sezione di Trieste, Italy}
\vskip 0.5 cm
\baselineskip=30pt
\centerline{\bf {\large Abstract.}}
We analize the algebraic structure of consistent and
covariant anomalies in gauge and gravitational theories: using a complex
extension of the Lie algebra it is possible to describe them in a unified
way. Then we study their representations by means of functional
determinants, showing how the algebraic solution determines the relevant
operators for the definition of the effective action. Particular attention
is devoted to the Lorentz anomaly: we obtain by functional methods the
covariant anomaly for the spin-current and for the energy-momentum tensor
in presence of a curved background. With regard to the consistent sector we
are able to give a general functional solution only for $d=2$: using the
characterization derived from the extended algebra, we find a continuous
family of operators whose determinant describes the effective action of
chiral spinors in curved space. We compute this action and we generalize
the result in presence of a $U(1)$ gauge connection.
\vskip 3.0 truecm
Accepted for publication in Fortschritte der Physik.
\end{titlepage}

\textwidth=15.5truecm
\oddsidemargin=0.5truecm
\evensidemargin=-0.09cm
\headsep=30pt
\topmargin=0.5truecm
\textheight=22.5truecm
\parindent=1.2truecm
\parskip=0.2truecm
\pagestyle{myheadings}
%\begin{document}
\setlength{\baselineskip}{12pt}

\newcommand{\EQ}{\begin{equation}}
\newcommand{\EN}{\end{equation}}
\newcommand{\BEF}{\begin{figure}}
\newcommand{\EF}{\end{figure}}
\newcommand{\bea}{\begin{eqnarray}}
\newcommand{\eea}{\end{eqnarray}}
\newcommand{\hs}{\hspace{0.1cm}}
\newcommand{\spz}{\hspace{0.7cm}}
\newcommand{\th}{\theta}
\newcommand{\s}{\sigma}
\newcommand{\goto}{\rightarrow}
\newcommand{\be}{\beta}
\newcommand{\zb}{\bar{z}}
\newcommand{\k}{\kappa}
\newcommand{\partiall}{\partial\hspace{-6pt}/}
\newcommand{\DI}{D\hspace{-8pt}/}
\newcommand{\inta}{\int d^{2n}x\sqrt{g}\,}
\newcommand{\ts}{\tilde{\sigma}}
\newcommand{\de}{\delta}
\newcommand{\lb}{\lambda}
\newcommand{\J}{\J_{\mu ab}}
\newcommand{\T}{\T^{\mu\nu}}
\newcommand{\sdag}{\,^{\dagger}}
\newcommand{\semi}{;\hfil\break}
\newcommand{\al}{\alpha}
\newcommand{\ua}{z_{1}}
\newcommand{\da}{z_{2}}

\section{Introduction}
\normalsize
In field theory anomalies appear as a breakdown at quantum level of
classical conservation laws: the symmetries of the original lagrangian
are not compatible, in general, with the quantum dynamics \cite{rfa}.
The breaking of a {\it global} symmetry (transformation laws  not
depending on space-time points) is not dangerous for the consistency of the
theory and sometimes is welcome: this breakdown is known
to lead to an understanding of $\pi^{0}$ decay and to a solution
of the $U(1)$ problem in Q.C.D. \cite{rfy}. On the contrary anomalies
affecting a {\it local} symmetry can modify crucially the Ward identities,
so that the properties of renormalizability and unitarity are lost.
In particular the perturbative consistency of a gauge theory is
destroyed in this case, giving a strong constraint on the possible
models of the fundamental interactions \cite{rfas}.
\newline
{}From a mathematical point of view we can consider the local anomalies
as {\it consistent anomalies}: they are
characterized by the fact that they satisfy an algebraic equation -- the
Wess-Zumino consistency condition -- derived by the group theoretical
properties of the classical symmetry. They split in two different types:
chiral anomalies (gauge, local Lorentz, diffeomorphism anomalies and their
supersymmetric version), which appear only in chirally asymmetric theories,
and conformal anomalies (conformal and superconformal anomalies) which
occur
in chiral and non chiral theories as well. A general feature of the
consistent anomaly is the loss of the classical tensorial properties of the
currents related to the broken symmetries: consistent anomalies do not
transform covariantly under the symmetry transformation of the classical
theory. With regard to  chiral theories we can define a second family,
the {\it covariant anomalies}: they are connected  with a redefinition of the
currents in such a way that the classical transformation laws are recovered.
\newline
{}From the physical point of view the two families are strongly distinguished
by
the fact that the consistent anomalies stem from dynamical currents,
coupled to the potentials of the theory, while the covariant ones do not.
In this sense the presence of consistent anomalies destroys the perturbative
consistency of the quantum theory: they entail a conflict between
renormalizability and unitarity.
\newline
When anomalies were discovered \cite{rfa} they rather appeared as a calculation
puzzle. In the first eighties it was realized that a rich algebraic and
geometrical structure subtends the existence of consistent anomaly: its
characterization as solution of a cohomological problem \cite{rfb} and its
relation with deep algebraic geometrical theorems, as the index theorem
\cite{rfc},
were crucial tools in understanding and solving many problems. In fact
these powerful techniques have been able to give general answers on
questions very difficult
to solve on the basis of the traditional perturbative methods \cite{rfac}.
\newline
Much less interest was devoted to study  on similar mathematical ground
the covariant anomaly,
that has not been considered as a fundamental object. Only recently,
after few years the seminal suggestions of \cite{rfd1,rfd2},
people realized that, even for
covariant anomalies,  a deep mathematical framework does exist describing their
structure \cite{rfe}.
\newline
A complementary approach to the problem of the anomaly relies on the
construction of some representations of the anomaly algebra: with regard
to theories
describing gauge or gravitational interaction of spin one-half fermions
this is obtained by means of the so called {\it fermionic determinant}.
In gauge theories different definitions of this object were given by the use of
$\zeta$-function technique \cite{rfat,rff}, finite mode regularization
\cite{rfg},
regularization of fermion propagator \cite{rfh}, reproducing the correct
perturbative results for the consistent anomalies. In the gravitational
case, at least to our knowledge, the only analytical definition and
explicit computation of the chiral determinant appeared in \cite{rfh,rfi},
claiming
the absence of diffeomorphism anomaly: no definition in term of $\zeta$-
function was given until now. In this framework covariant anomalies
sporadically appeared as a mistake in the regularization procedure
\cite{rfl}: a
first step towards a systematic construction of a functional representing the
algebra of covariant anomalies has been presented in \cite{rfm}.
\newline
In this paper we generalize
those results for a gauge theory to an arbitrary even dimension:
consistent and covariant gauge anomalies are different solutions of an extended
cohomological problem (Sect.2) and they can be obtained from different
fermionic determinants (Sect.3).
\newline
Then we try to generalize these results to the
gravitational case (Sect.4): the situation is subtler due to the difficult
use of the $\zeta$-function technique. We are able to solve completely the
covariant problem and, using the properties of the extended algebra, we
find the correct operator whose $\zeta$-function determinant gives the
consistent anomaly at $d=2$ (Sect.5). We perform the calculation of this
determinant, even in presence of a $U(1)$ gauge connection (Sect.6),
that turns out to agree with Leutwyler's result \cite{rfi},
obtained in a more indirect way.
Our procedure represents also a manifestly diffeomorphism-invariant
calculation of the
determinant (in \cite{rfi} polynomial counterterms are needed to achieve
the general covariance).
\newline
We do not consider here the case of non trivial fiber bundles (non trivial
principal bundles for gauge theories and non parallelizable manifold for
spinors in curved space),  but we hope in future to extend our results to
this interesting situation: anyway some hints are given towards it. We
conclude remarking that there is an increasing physical interest for the
study of
the covariant structure of the anomalies: in a recent paper \cite{rfn} strong
relations between quantization of anomalous theories and covariant
currents have been discovered.
Besides, we have found stimulating links between Chern-Simons
theory in $2n-1$ dimensions and covariant anomalies in $2n-2$
dimensions presented in \cite{rfo}.
The complexification of the gauge group, that is the main tool in
our description of covariant anomalies, has also been used, in a different
context, by Witten \cite{rfp}.
\section{Algebraic structure of gauge anomalies}
\normalsize
In this section we briefly review the definitions of consistent and
covariant anomaly for a gauge theory in flat d-dimensional euclidean
space-time. We recall the algebraic description of the consistent anomaly and
we show how to extend this characterization to the covariant one.
\newline
\subsection{Consistent anomalies}
Let $\Gamma$[A] be the vacuum functional in presence of the external gauge
field $A=A_{\mu}^a T_a dx^{\mu}$ and $J_{\mu}^a(x)=\frac{\delta\Gamma[A]}
{\delta A^{\mu}_a}(x)$ the gauge field current. In presence of chiral
fermions  the functional $\Gamma[A]$ is not gauge invariant.
If\[\delta_{\lambda} A_{\mu}=D_{\mu}\lambda=\partial_{\mu}\lambda +
[A_{\mu},\lambda]\]describes the infinitesimal gauge variation due to the Lie
algebra valued parameter $\lambda=\lambda_{a} T_{a}$ and
\[\delta(\lambda)=
\int d^{2n-2}x (\delta_{\lambda} A_{\mu}(x))_{a}\frac{\delta}{\delta A_{\mu}
^{a}  (x)}\]
is the operator realizing the transformation on some functional of
$A_{\mu}(x)$ , the symmetry breaking manifests itself by the occurrence of an
(integrated) anomaly $\Delta(\lambda,A)$
\begin{eqnarray}
&\delta(\lambda)\Gamma[A] = \Delta(\lambda,A)=\int d^{2n-2}x Tr[\lambda G](x),
\nonumber\\
&D_{\mu}J^{\mu}_{a} = -G_{a}[A].
\end{eqnarray}
The use of $2n-2$ as euclidean dimensions will become clear in the following
discussions.
\newline
Obviously $\delta(\lambda)$ represents the Lie algebra
\begin{equation} [\delta(\lambda_{1}),\delta(\lambda_{2})]=\delta([\lambda_
{1},\lambda_{2}]), \end{equation}
forcing the Wess-Zumino (W-Z) consistency condition upon $\Delta(\lambda,A)$
\begin{equation}\delta(\lambda_{1})\Delta(\lambda_{2},A)-\delta(\lambda_{2})
\Delta(\lambda_{1},A)=\Delta([\lambda_{1},\lambda_{2}],A).\end{equation}
Any local functional of $A_{\mu}$, linear in $\lambda$, which satisfies
(3) is called consistent anomaly. One can put (3) in a more compact form
by turning $\lambda$ into a ghost field $c$: the W-Z condition is equivalent
to
\begin{equation}
s\Delta(c,A)=0,
\end{equation}
with the definition for $s$
\begin{eqnarray}
&sA=-Dc,\nonumber\\
&sc=-\frac{1}{2}[c,c],\nonumber\\
&s^{2}=0;
\end{eqnarray}
$s$ is the B.R.S.T. operator, characterizing the anomaly as solution of a
cohomological problem. Defining the  field strength \[F=dA+AA=\frac{1}{2}F_{
\mu\nu}dx^{\mu}dx^{\nu},\]
one can prove the ``Russian formula''
\begin{equation}
\tilde{F} =F,
\end{equation}
where
\begin{eqnarray}
&\tilde{F}=\tilde{d}\tilde{A}+\tilde{A}\tilde{A},\nonumber\\
&\tilde{d}=d+s,\nonumber\\
&\tilde{A}=A+c.\nonumber\\
\end{eqnarray}
In the case of a trivial bundle (I recall that the relevant structure for
gauge theory is a principal bundle $P(M,G)$ where $M$ is the compactified
euclidean space-time and $G$ is the Lie group underlying the theory) a
solution of (4) is easily given applying simultaneously the ``Russian
formula'' and the ``transgression formula'' \cite{rfq}
\begin{equation}
P(F^{n})=d\omega^{0}_{2n-1}(A),
\end{equation}
$P$ being a symmetric polynomial on the Lie algebra, invariant under the
adjoint action of the group, and $\omega^{0}_{2n-1}(A)$ the Chern-Simons
form.
\newline
Starting from the obvious identity
\begin{equation}
P(\tilde{F}^{n})=\tilde{d}\omega_{2n-1}(\tilde{A}),
\end{equation}
we expand the (2n-1) form $\omega_{2n-1}(\tilde{A})$ in powers of $c$
\begin{equation}
\omega_{2n-1}(\tilde{A})=\sum_{k=0}^{2n-1} \omega_{2n-1-k}^{k}(A,c),
\end{equation}
the lower index giving the form degree and the upper one the ghost number.
The so-called ``descent equations'' are obtained matching the powers of $c$
\begin{eqnarray}
&d\omega^{0}_{2n-1}(A)=P(F_{n}),\nonumber\\
&s\omega^{k}_{2n-1-k}(A,c)+d\omega^{k+1}_{2n-2-k}(A,c)=0.
\end{eqnarray}
For $k=1$ the integration gives
\begin{equation}
s\int\omega^{1}_{2n-2}(A,c)=0,
\end{equation}identifying $\Delta(A,c)=\int\omega_{2n-2}^{1}(A,c)$ as a good
 candidate for  the consistent anomaly.
An explicit expression is
\begin{equation}
\Delta(c,A)=\int\omega^{1}_{2n-2}(A,c)=
n(n-1)\int_{0}^{1}dt\, P(dc+[A,c],tA,F^{n-2}_{t}), \label{eq:ras}
\end{equation}
with $F_{t}=t\,dA+t^{2}AA$.
\newline
We remark that the W-Z condition is linear in $\Delta$; hence, in order to find
the correct coefficient we need more informations about the matter content
of the theory, or, equivalently, about the definition of $\Gamma[A]$. We
remember that one can always add to $\Delta$ a term of the type $s\alpha(A)+
d\beta(A,c)$ with the correct canonical dimensions: this freedom corresponds
to the choice of the representative in the cohomology class.
\newline
For a non-trivial principal bundle the situation is
slightly more complicated: a good solution
can be obtained \cite{rfd1} fixing a background connection $A_{0}$,
belonging to the same bundle of $A$. The formula (1.7) is modified as follows
\begin{equation}
P(F^{n})-P(F^{n}_{0})=d\omega^{0}_{2n-1}(A,A_{0}),
\end{equation}
showing that $P(F^{n})$ is not exact as in the case of a trivial bundle:
actually we consider $P(F^{n})$ as a form on the base manifold, the
projection of a (exact) form defined on the whole principal bundle. The
anomaly acquires a dependence on $A_{0}$: it corresponds, as we will see, to an
equivalent obstruction in defining a Weyl determinant in this non
topologically trivial situation.
\subsection{Covariant anomalies}
One can easily show that if the consistent anomaly is different from zero
the current $J^{\mu}_{a}(x)$ does not transform covariantly under gauge
transformations. In general it is possible to find a polynomial $X^{\mu}_{a}
(A)$ which makes the current covariant \cite{rfr}. One defines a new current
\begin{equation}
\hat{J}^{\mu}_{a}=J^{\mu}_{a}+X^{\mu}_{a},
\end{equation}
with the property
\begin{equation}
\delta(\lambda)\hat{J}_{\mu}=-[\lambda,\hat{J}_{\mu}],
\end{equation}
thereby recovering the classical tensorial transformation.
The polynomial $X^{\mu}_{a}(A)$ is called the Bardeen-Zumino
counterterm and the covariant
divergence of $J^{\mu}_{a}$ is known as the covariant anomaly:
\begin{equation}
D_{\mu}\hat{J}^{\mu}_{a}=-\hat{G}_{a}(A).
\end{equation}
In order to understand this operation two remarks are useful: firstly the
redefinition of $J^{\mu}_{a}$ does not correspond to an allowed
one of the vacuum functional. In other words $\hat{G}_{a}(A)$ does
not belong to the same cohomological class of $G_{a}(A)$ (and it does not
satisfy the W-Z condition). Secondly, even in the abelian case there is a
difference between covariant and consistent anomaly: the functional form is
equal but the numerical coefficient is different.
\newline
In $d=2n-2$ space-time dimensions the explicit expression for $\hat{G}(A)$ is
known \cite{rfr} (for any simple group)
\begin{equation}
\Delta_{cov}(A,c)=\int dx^{2n-2} Tr[c\,\hat{G}]=n\int P[c\,,F^{n-1}].
\end{equation}
Anyway $\Delta_{cov}$ can also be algebrically characterized, embedding it
into the solution of a cohomological problem \cite{rfd1}.
\newline
We make the hypothesis
that there is a subgroup $K$ of a Lie group $\tilde{G}$ with the
properties that the invariant symmetric polynomial $P$ vanishes when its
arguments are restricted to $Lie K$ (we assume again that $P(M,\hat{G})$ is
trivial). It is possible to decompose $A$ and $c$ along $Lie K$ and an
invariantly defined orthogonal complement $(Lie K)_{\perp}$
\begin{eqnarray}
&A=A_{K}+A_{\perp},\nonumber\\
&c=c_{K}+c_{\perp},
\end{eqnarray}
with $A_{K},c_{K}\in Lie K$ and $A_{\perp},c_{\perp}\in (Lie K)_{\perp}$
We remember that $(Lie K)_{\perp}$ is not in general a Lie algebra.
\newline
The consistent anomaly $\Delta(c,A)$, as it stands, does not vanish along
$Lie K$ but it reduces to $\Delta(c_{K},A_{\perp})$: in reference
\cite{rfd1}  the existence was proved of a general counterterm
$\Gamma_{B} [A]=\int X(A_{\perp},A_{K})$,
local polynomial in $A_{\perp}$ and $A_{K}$, called the Bardeen
counterterm, that, added to the vacuum functional, gives
\[\hat{\Delta}(c_{\perp},A_{\perp},A_{K})=\Delta(c,A)+s\,\Gamma_{B}[A],\]
\[s\hat{\Delta}(c_{\perp},A_{\perp},A_{K})=0,\]
\[\hat{\Delta}(c_{\perp},A_{\perp},A_{K})=n\int\int_{0}^{1}dt P(c_{\perp},F^
{n-1}(A_{K}+tA_{\perp})+n(n-1)\]
\EQ
\int\int_{0}^{1}dt P(A_{\perp},\,t^{2}[A_{\perp},c_{\perp}]-t[A_{\perp}
-c_{\perp}]_{\perp}-[A_{\perp},c_{\perp}]_{K},\,F^{n-2}(A_{K}+tA_{\perp})).
\EN
Projecting $\hat{\Delta}(c_{\perp},A_{\perp},A_{K})$ on $A_{\perp}=0$
\begin{equation}
\hat{\Delta}(c_{\perp},0,A_{K})=n\int P(c_{\perp},F^{n-1}(A_{K}).
\label{eq:coo}
\end{equation}
The comparison with the consistent solution
\begin{equation}
\int\omega^{1}_{2n-2}=\int P(c,F^{n-1})+...
\end{equation}
shows the appearance of the factor $n$ \cite{rfr}. Now if $K$ is the
structure group $G$, $\hat{\Delta}(c_{\perp},0,F^{n-1})$ is the covariant
anomaly. Hence, taking a suitable embedding of $G$
in some larger group $\tilde{G}$, it is possible
to obtain a solution of the cohomological problem (4) that reduces
to the covariant anomaly after a projection. The same is true also in
presence of a non trivial $P(M,\tilde{G})$, requiring that the bundle is
reducible to $P(M,G)$. An interesting feature is that, in this case, the
covariant anomaly, at variance with the consistent one, does not depend on
some fixed background connection: we will discuss this difference in the
framework of the functional approach.
All these properties have been further explored, and new descent equations,
describing the covariant anomalies, have been obtained using concepts like
vertical cohomology and local B.R.S.T. symmetry \cite{rfe}.
For our purposes the
characterization we have just described is sufficient: our choice of the group
in which to embed $G$ is different from the original proposal \cite{rfd1}
and it was firstly discussed in \cite{rfm}.
As we will see later on, it is directly related to a
functional approach and it is also possible to obtain the consistent
anomaly. We will choose $G=SU(N)$.
\subsection{The complex extension}
Let us take complex values for the gauge potentials
\begin{equation}
A=A_{a}T_{a}\Longrightarrow \,\,
\hat{A}=(A^{1}_{a}+iA^{2}_{a})T_{a}=\hat{A}_{a}
\hat{T}_{a},
\end{equation}
\begin{equation}
\hat{T}_{a}=(T_{a},iT_{a})
\end{equation}
where now $\hat{T}_{a}$ is a basis for $SL(N,C)$:
applying the previously derived formalism we identify
\begin{eqnarray}
&\tilde{G}=SL(N,C) & A_{K}=A^1_{a}T_{a}=A_{1},\nonumber\\
&K=SU(N)            & A_{\perp}=A^{2}_{a}iT_{a}=A_{2},
\end{eqnarray}
with $c_{k}=c_{1}$ and $c_{\perp}=c_{2}$.
\newline
Then we have to exhibit an invariant symmetric polynomial on $SL(N,C)$
vanishing on $SU(N)$: the correct choice is
\begin{equation}
P(A_{1},A_{2})=\frac{1}{2i}[P(F^{n}(\hat{A})-P^{\ast}(F^{n}(\hat{A})].
\end{equation}
We can specialize equation (20) with the result
\[s\tilde{\Delta}(A_{1},A_{2};c_{2})=0,\]
\[\tilde{\Delta}_{cov}(A_{1},A_{2};c)=\frac{n}{2i}\int_{0}^{1}dt\int P(c_{2},
\, F^{n-1}(A_{1}+tA_{2}))+\]
\[+\frac{n(n-1)}{2i}\int_{0}^{1}dt\int P(A_{2},(t^{2}-1)\, [A_{2},c_{2}],\,
F^{n-2}(A_{1}+tA_{2}))-\]
\EQ
-(c_{2}\rightarrow-c_{2} \, ; \, A_{2}\rightarrow-A_{2}).
\EN
For $A_{2}=0$ (real projection), with the choice $c_{2}=i\hat{c}_{2}$ we
obtain
\begin{equation}
\Delta_{cov}(A_{1},0,c_{2})=n\int P(\hat{c}_{2},F^{n-1}(A_{1})).
\end{equation}
But we can derive from the present formalism also the usual consistent
anomaly: if we do not require the vanishing of the symmetric invariant
polynomial on $SU(N)$ and we choose
\[P(A_{1},A_{2})=P(F^{n}(A))\]
or
\[P(A_{1},A_{2})=P^{\ast}(F^{n}(A))\]
we obtain a solution of the cohomological problem under the form
(actually * means complex conjugation):
\[\tilde{\Delta}_{con}(A_{1},A_{2},c_{1},c_{2})=\tilde{\Delta}^{1}_{con}=
\int\omega^{1}_{2n-2}(A,c)\]
or
\[\tilde{\Delta}_{con}(A_{1},-A_{2},c_{1},-c_{2})=\tilde{\Delta}^{2}_{con}=
\int\omega^{1}_{2n-2}(A^{\dagger},c^{\dagger}).\]
It is rather clear that the projection on $SU(N) \,\, (c_{2}=0)$ reproduces,
after the limit $A_{2}=0$, the usual expression for the consistent anomaly
$\Delta_{con}(A_{1},c_{1})$.
This is not really the more general way to embed the consistent solution
into the extended problem, although it is the basic one: we are going to see
that any solution that reduces to the consistent one on $SU(N)$ is
built in terms of a covariant solution and
$\tilde{\Delta}_{con}(A_{1},A_{2};c_{1},c_{2})$ we have just found.
\newline
Any invariant polynomial
\begin{equation}
\alpha P(F^{n}(A))+\beta P^{\ast}(F^{n}(A))
\end{equation}
with $\alpha +\beta =1$ reduces to the usual $SU(N)$ polynomial $P(F^{n}(
A_{1}))$ on $A_{2}=0$; conversely, applying the descent equation to
(29),
one recovers a solution
$\tilde{\Delta}^{\alpha ,\beta}(A_{1},A_{2};c_{1},c_{2})$
that gives
\begin{equation}
\tilde{\Delta}^{\alpha ,\beta}(A_{1},0;c_{1},0)=\Delta_{con}(A_{1},c_{1}).
\end{equation}
Let us see how $\tilde{\Delta}^{\alpha ,\beta}$ is constructed: we rewrite
(29) using $\beta =1-\alpha$
\begin{equation}
2\alpha\,[\frac{1}{2}(P(F^{n})-P^{\ast}(F^{n}))]+P(F^{n})
\end{equation}
and immediately get
\begin{equation}
\tilde{\Delta}^{\alpha ,\beta}=2\alpha\tilde{\Delta}_{cov}(A_{1},A_{2},c_{2})
-2\alpha s\Gamma_{B}(A_{1},A_{2})+\tilde{\Delta}^{1}_{con}(A_{1},-A_{2};c_{1},-
c_2);
\end{equation}
$\tilde{\Delta}^{\alpha ,\beta}$ appears (up to coboundary terms) to be the
sum of a covariant like solution $2\alpha\tilde{\Delta}_{cov}$ (vanishing
on $SU(N)$) and the consistent solution $\tilde{\Delta}^{1}_{con}$: in other
words two solutions of the extended cohomological problem that reduce to
the usual consistent one on $SU(N)$ (and $A_{2}=0$) differ by a term
proportional to the covariant solution. All the informations about the
consistent anomaly is encoded in $\Delta^{1}_{con}$ or $\Delta^{2}_{con}$.
The natural question that now arises is: in the extended situation
\[s\tilde{\Delta}_{cov}=0,\]
\[s\tilde{\Delta}^i_{con}=0,\]
could $\tilde{\Delta}_{cov}$ and $\tilde{\Delta}_{con}^{i}$ differ by some
term of the type $sH(A)$?
\newline
One can prove with some explicit examples (for instance in the simplest case
$d=2$) that the answer is negative. The two solutions belong to different
cohomology classes of the $s$ operator defined on $SL(N,C)$.
\newline
Before closing this section we rewrite the equation for covariant and
consistent anomalies, embedded on $SL(N,C)$, in a more handful way for the
future functional applications. Coming back to the original Ward operator
$\delta(\lambda)$ for $SL(N,C)$, $\lambda=\lambda_{a}\hat{T}_{a}$ we write
\begin{equation}
\delta(\lambda)=\delta_{1}(\lambda_{1})+\delta_{2}(\lambda_{2})
\end{equation}
with $\delta_{1}(\lambda_{1})$ generating gauge transformations of $SU(N)$
type and $\delta_{2}(\lambda_{2})$ acting on the orthogonal invariant
complement. The algebra of $\delta_{1}$ and $\delta_{2}$ is
\begin{eqnarray}
&[\delta_{1}(\lambda_{1}),\delta_{1}(\lambda_{2})]=\delta_{1}([\lambda_{1},
\lambda_{2}]),\nonumber\\
&[\delta_{2}(\lambda_{1},\delta_{2}(\lambda_{2})]=-\delta_{1}([\lambda_{1},
\lambda_{2}]),\nonumber\\
&[\delta_{1}(\lambda_{1}),\delta_{2}(\lambda_{2})]=\delta_{2}([\lambda_{1},
\lambda_{2}]),    \label{eq:grig}
\end{eqnarray}
leading to the W-Z conditions
\begin{eqnarray}
&\delta_{1}(\lambda_{1})a_{1}(\lambda_{2})-\delta_{1}(\lambda_{2})a_{1}(
\lambda_{1})=a_{1}([\lambda_{1},\lambda_{2}]),\nonumber\\
&\delta_{2}(\lambda_{1})a_{2}(\lambda_{2})-\delta_{2}(\lambda_{2})a_{2}(
\lambda_{1})=-a_{1}([\lambda_{1},\lambda_{2}]),\nonumber\\
&\delta_{1}(\lambda_{1})a_{2}(\lambda_{2})-\delta_{2}(\lambda_{2})a_{1}
(\lambda_{1})=a_{2}([\lambda_{1},\lambda_{2}]),
\end{eqnarray}
where $a_{1}=\delta_{1}\Gamma$ and $a_2=\delta_{2}\Gamma$.
\newline
The covariant solution corresponds to
\begin{eqnarray}
&a_{1}=0,\nonumber\\
&a_{2}\neq 0.   \label{eq:cip}
\end{eqnarray}
The consistent one, as it was shown in reference [16], is
\begin{equation}
a_{2}=\pm ia_{1}.  \label{eq:ciop}
\end{equation}
The sign relies on the choice of $P$ or $P^{\ast}$ as invariant polynomial.
Obviously this is not the only way to characterize the consistent solution:
$a_{2}$ can always differ, as we have seen, by a term proportional to the
covariant anomaly
\begin{equation}
\bar{a}_{2}=\pm ia_{1}+\alpha \, a^{cov}_{2}.
\end{equation}
So it is sufficient to look for solutions respecting (37): all the other ones
are obtained by adding to the functional $\Gamma$ a functional
$\alpha\Gamma_{cov}$ with
\[\delta_{1}\Gamma_{cov}=0,\]
\[\delta_{2}\Gamma_{cov}=a^{cov}_{2}.\]
At the end we remark that (36) and (37) are particular choices into
the cohomology classes giving, in the limit $A_{2}=0$, the canonical form for
the anomalies: hence the equality in eq.(37) is to be understood modulo
coboundary terms (of $SL(N,C)$): the projection on $A_{2}=0$ produces the
correct anomaly (on $SU(N)$) only for representatives satisfying exactly
(36), (37).
\vfill\eject

\section{Algebras and functional determinants in gauge theories}
In the previous sections we did not care about the definition of the vacuum
functional $\Gamma [A]$: it is to be obtained by some regularization
procedure from the formal expression
\[\Gamma [A]=-\ln \int [D\phi]\exp[-S_{cl}(A)],\]
$S_{cl}(A)$ being the classical action for the field $\phi$ coupled to an
external gauge field $A$.
\newline
Perturbation theory induces the fact that two different regularization
procedures on $\Gamma [A]$ are distinguished only by some local term in the
field $A$ and its derivatives, of the correct canonical dimensions. To choose
a different regularization corresponds, for the anomaly, to take a
different representative in the cohomology class. If we extend the gauge
group we have to find a definition for $\Gamma [\hat{A}]$: but now there are
two non trivial independent classes of cohomology for the extended problem.
As matter of fact we can construct two non trivial
representations of the algebra (\ref{eq:grig}), giving respectively the
different solutions, characterized by the conditions (\ref{eq:cip}) and
(\ref{eq:ciop}). They are not distinguished by coboundary terms, so that
only the second solution, giving the consistent anomaly, turns out to
lead to the correct functional:
eq.(\ref{eq:ciop}) characterizes the correct vacuum functional for an
anomalous $SU(N)$ gauge theory. We will use this property also
in the gravitational case in order to find the effective action at $d=2$.
\newline
\subsection{The problem of Weyl determinant}
The classical action for spinors coupled to a gauge or a gravitational
background is
\begin{equation}
S_{cl}=\int d^{2n}\Omega \, \bar{\psi}\DI\, \psi
\end{equation}
with $\DI$ some first order operator of Dirac type and $d^{2n}\Omega$ an
appropriate measure.

Formally it results
\begin{equation}
\Gamma[A]=-\ln Det \DI\, (A).
\end{equation}
Immediately we are faced with a basic difficulty in defining the determinant
for chiral fermions: in this case $\DI$ is the {\it Weyl operator}
that maps a chiral spinor on a spinor of opposite chirality
\begin{equation}
\DI : \Gamma (S_{+})_{M} \rightarrow \Gamma (S_{-})_{M}
\end{equation}
where $\Gamma(S_{+})_{M} \, (\Gamma(S_{-})_{M})$ is the space of right (left)
sections of the vector bundle associated by the Dirac representation to the
Spin-principal bundle on $M$. $\Gamma (S_{+})_{M}$ and $\Gamma (S_{-})_{M}$ are
different Hilbert spaces and there is no canonical isomorphism
between them: $\DI$ does not map a Hilbert space into itself and we have
no canonical way to define a meaningful eigenvalue problem for this
operator. A way out is to modify the Weyl operator in order to have a good
eigenvalue problem but, in general, some of the classical properties are
lost after the modification. The Dirac operator has no problem
\begin{eqnarray}
&\DI_{D} : \Gamma (S)_{M} \rightarrow \Gamma (S)_{M}\nonumber\\
&\Gamma (S)_{M}=\Gamma (S_{+})_{M}\oplus\Gamma (S_{-})_{M}.
\end{eqnarray}
The eigenvalue problem is meaningful
\[\DI_{D}\Psi_{n}=\lambda_{n}\Psi_{n}\]
$n$ is an integer (we suppose to work on a compact manifold).
A gauge transformation on $\DI_{D}$ acts as
\begin{equation}
g : \DI_{D} \rightarrow U^{-1}\DI_{D}U.  \label{eq:ciccio}
\end{equation}
The covariant transformation (\ref{eq:ciccio}) implies that the eigenvalues
are constant under the gauge action.
In general $|\lambda_{n}|$ grows with $n$ so the naive definition
\begin{equation}
Det \DI_{D}=\prod_{n}\lambda_{n}
\end{equation}
is meaningless. We can use an analytic regularization, the $\zeta$-function
regularization \cite{rfat}, that does not change the eigenvalues but simply
damps the
contribution of the highest ones. The so defined determinant is invariant
under gauge transformations. In the Weyl case one immediately recognizes that
even if the covariance is conserved by the original operator, the same may
not be true for the modified one. Let us see what happens if we try to
apply directly the $\zeta$-function machinery to the Weyl operator: we fail
because there is no way to define its complex power \cite{rfs}.
As an example we consider the $d=2$ flat case, where the operator is
\begin{equation}
\DI =i\sigma_{\mu}( \partial_{\mu}+A_{\mu} ),
\end{equation}
$A_{\mu}$ being an antihermitian potential and $\sigma_{\mu}$ one of the
two inequivalent representations of the Weyl algebra \cite{rft}. The principal
symbol \cite{rfs} of $\DI$ is
\begin{equation}
a_{1}(x,\xi)=-\sigma_{\mu}\xi_{\mu}=-\xi_{1}-i\xi_{2}
\,\,\,\,\,\,\,\,\,\,\,\,\,\,  \xi
\in R,
\label{eq:alba}
\end{equation}
where we have chosen $\sigma_{1}=1$ and $\sigma_{2}=i$. To define a complex
power of $\DI$ we need the existence of a ray of minimal growth \cite{rfs},
namely
we have to find an angle $\phi$ in the complex plane of the variable
$\lambda=t\exp i\phi$
such that
\begin{equation}
t\exp i\phi - a_{1}\neq 0\hspace{7 mm}\forall t>0, \, \, |\xi|=1\,\,.
\end{equation}
This is clearly incompatible with the expression (\ref{eq:alba}). We see that
already for a gauge theory on a flat manifold the Weyl determinant cannot be
defined, at least by the $\zeta$-function method. The same is true in
curved space. The Dirac case is different: the principal symbol is
\[a_{1}(x,\xi)=-\gamma_{\mu}\xi_{\mu}\]
that does possess the ray ($\gamma_{\mu}$ are the Dirac matrices).
\subsection{The definition of Weyl determinants}
In order to obtain a good eigenvalues problem it is quite natural to fix an
isomorphism $T$
\begin{equation}
T:\Gamma(S_{-})\rightarrow\Gamma(S_{+})
\end{equation}
and to define
\begin{equation}
Det \DI\equiv Det(T\DI )
\end{equation}
where now the operator $T\DI$ admits a well defined eigenvalues problem
even if covariance is, a priori, lost: our task is to find such a $T$ to
recover, under a gauge variation, the consistent anomaly from the just
defined functional. In the gauge case the question is easily solved, also
for non-trivial principal bundles, and the embedding
into a $SL(N,C)$ theory determines not only the consistent but even the
covariant solution of the extended problem. The physical intuition gives
immediately a result: in the case of a trivial bundle one can always describe
a right fermion interacting with a gauge field as a Dirac fermion in which
only the right part is coupled with the gauge field. In $d=2n$ the Weyl
operator is
\[\DI =i\sigma_{\mu}( \partial_{\mu}+A_{\mu} ),\]
$\sigma_{\mu}$ are the Weyl matrices obeying
\begin{equation}
\sigma_{\mu}\tilde{\sigma}_{\nu}+\sigma_{\nu}\tilde{\sigma}_{\mu}=
2\delta_{\mu\nu},
\end{equation}
$\tilde{\sigma}_{\mu}$ is the other inequivalent representation of the
algebra, existing for $d=2n$ \cite{rft}. We can use
\begin{equation}
\hat{\DI}=i\gamma_{\mu}( \partial_{\mu}+(\frac{1+\gamma_{2n+1}}{2
})A_{\mu})
\end{equation}
$\gamma_{\mu}$ being the Dirac matrices. $\hat{\DI}$ has a good
eigenvalues problem and admits $\zeta$-function regularization possessing
a correct principal symbol: obviously in this case the covariance is lost.
It happens that
\begin{equation}
\hat{\DI}=\left( \begin{array}{cc}0&\DI\\
i\tilde{\sigma}_{\mu}\partial_{\mu}&0
                     \end{array}
             \right),
\end{equation}
so that the determinant is
\begin{equation}
Det \hat{\DI}=Det[(i\tilde{\sigma}_{\mu}\partial_{\mu})\DI\,].
\end{equation}
The isomorphism $T$ appears to be the free kinetic Weyl operator of opposite
chirality. In some sense it corresponds to a particular normalization of
the determinant: $i\tilde{\sigma}_{\mu}\partial_{\mu}$ has no dynamical
contribution and we can formally subtract it.
\newline
{}From a more formal point of view, we require (always for a trivial bundle):
\begin{enumerate}
\item $Det(T\DI)$ gives rise to the consistent anomaly (solution
(\ref{eq:ciop}) of the extended W-Z).
\item The anomaly should be local in the gauge field.
\item $Det(T\DI)$ is smoothly connected to the free case.
\item $Det(T\DI)$ is a functional only of the gauge field $A_{\mu}$ (for non
trivial $P(M,G)$ this is not possible as we will see).
\end{enumerate}
Essentially we are trying to define the determinant of a first order operator
in terms of the determinant of a second order one. Moreover we extend the
gauge group to explore the possible different representation of the
extended anomaly algebra, varying the isomorphism $T$.
\newline
A good general candidate for $T$ is
\begin{equation}
T(r)=\tilde{\DI}\,(r)=i\tilde{\sigma}_{\mu}(\partial_{\mu}+r \hat{A}_{\mu}
^{\dagger})\hspace{.5in}r\in\bf{R}. \label{eq:pippo}
\end{equation}
Now some remarks are in order.
We can always take the matrices $\tilde{\sigma}_{\mu}$ as
${\sigma}_{\mu}^{\dagger}$, from the general properties of the Weyl
algebra, thereby getting for $r=1$:
\[\tilde{\DI}= {\DI}\,^{\dagger}\]
and:
\begin{equation}
Det(T(1)\DI)=Det({\DI}\,^{\dagger}\DI)= Det|(\DI)|^2 \,.
\end{equation}
For $r=1$ we lose the phase of the determinant on which the {\it unextended}
anomaly relies [24]: it is well known that the modulus of the Weyl
determinant is equal (apart from some regularization terms) to the Dirac
determinant.
\newline
Let us study the determinant (\ref{eq:pippo}): conditions 3. and 4. are
obviously satisfied by the $\zeta$-function definition. To test different
$r$'s we perform the variations $\delta_1(\al)$ and $\delta_2(\al)$ on
$\hat{A}_{\mu}$
and $\hat{A}_{\mu}^{\dagger}$: we compute the answers of the determinant
we have hitherto introduced and try to find the values of $r$ for which
\begin{equation}
a_1(\al)=\pm ia_2(\al)
\end{equation}
with
\[a_1(\al)=\delta_1(\al)\ln Det (T(r) \DI)\]
\[a_2(\al)=\delta_2(\al)\ln Det (T(r) \DI)\]
where $a_1$ and $a_2$ are local on $\hat{A}$ and $\hat{A}^{\dagger}$. We
shall also find a $T(r)$ for which
\begin{equation}
a_1=0\hspace{.5in}a_2\neq 0.
\end{equation}
Using standard $\zeta$-function formalism \cite{rff} we obtain
\begin{equation}
\Gamma[\hat{A},\hat{A}^{\dagger};r]=\frac{1}{F(r)}\frac{d}{ds} Tr[(\tilde{\DI
}\,(r)\DI\,)^{-s}]_{s=0}
\end{equation}
$Tr$ is an operatorial trace and $F(r)$ is some normalization factor that
gives to the effective action its correct weight.
\newline
After some calculations we find:
\[\delta_1(\alpha)\Gamma\!=\frac{1}{F(r)}\frac{i}{(4\pi)^n}\int d^{2n}x \,
Tr[H_n(r)(1-r)\alpha-\tilde{H}_n(r)(1-r)\alpha]-\]
\EQ
-\frac{1}{F(r)}(r-1)r \, \frac{d}{ds}[\,s\,Tr\{\tilde{\sigma}_{\mu}
[\hat{A}^{\dagger}_{\mu}, i\alpha]\DI\,(\tilde{\DI}\,(r)\DI\,)^{-s-1}\}]
_{s=0}
\EN
and
\[\delta_2(\alpha)\Gamma\!=\frac{1}{F(r)}\frac{1}{(4\pi)^n}\int d^{2n}x \,
Tr[H_n(r)(1+r)\alpha+\tilde{H}_n(r)(1+r)\alpha]-\]
\EQ
-\frac{1}{F(r)}(r-1)r \, \frac{d}{ds}[\,s\,Tr\{\tilde{\sigma}_{\mu}
[\hat{A}^{\dagger}_{\mu}, \alpha]\DI\,(\tilde{\DI}\,(r)\DI\,)^{-s-1}\}]
_{s=0}\,\, ,
\EN
where
\[H_n=H_n(\tilde{\DI}\,(r)\DI\,),\]
\[\tilde{H}_n=H_n(\DI\,\tilde{\DI}\,(r)\,),\]
$H_n$ is the n-th coefficient of the heat kernel expansion of the operator
in question \cite{rfs}.
\newline
The second term in the variation is in general non local requiring the use
of the inverse of a differential operator: only in the abelian case, where
the commutators disappear, we have a local variation for any value of $r$.
The non local term disappears for $r=0$ and $r=1$.
\newline
\underline{{\bf $r=0$ (the consistent case)}}:
\[a_1(\alpha)=\frac{1}{F(0)}\frac{i}{(4\pi)^n}\int d^{2n}x \,
Tr[(-H_n(0)+\tilde{H}_n(0))\alpha],\]
\EQ
a_2(\alpha)=\frac{1}{F(0)}\frac{1}{(4\pi)^n}\int d^{2n}x \,
Tr[(-H_n(0)+\tilde{H}_n(0))\alpha].
\EN
So $a_2=-ia_1$. Therefore $r=0$ gives the consistent anomaly; the correct
choice for $F(0)$ is $1$. We note that in this case the relevant operator is
\[ (i\tilde{\sigma}_{\mu}\partial_{\mu})\DI\Rightarrow T(0)=
i\tilde{\sigma}_{\mu}\partial_{\mu}\]
as we have guessed from physical arguments. The usual determinant is
obtained by projecting $A_2=0$.
\newline
\underline{{\bf $r=1$ (the covariant case)}}:
$$
a_1(\alpha)=0,
$$
\EQ
a_2(\alpha)=\frac{1}{F(1)}\frac{1}{(4\pi)^n}\int d^{2n}x
\,2\,Tr[(-H_n(1)+\tilde{H}_n(1))\alpha],
\EN
$a_1(\alpha)\!=\!0$ and $a_2(\alpha)\!\neq\!0$, namely the covariant solution.
Taking $F(1)\!=\!2$ one can compute in the limit $A_2=0$ the trace
\cite{rft}: the
result is
\begin{equation}
\frac{1}{(4\pi)^n}\frac{sgn(\sigma)}{n!}\int d^{2n}x \,
\epsilon_{\mu_{1}...\mu_{2n}} Tr[F_{\mu_{1}\mu_{2}}...F_{\mu_{2n-1}\mu_{2n}}
\alpha],
\end{equation}
which corresponds to the cohomological result (\ref{eq:coo}), taking into
account the correct normalization factors for the polynomial $P$. We remark
that the covariant anomaly can be, in this way, obtained starting from
$A_2=0$, varying the modulus of the Weyl determinant with a general
transformation of $SL(N,C)$. For general $r$ we have again solutions of
the cohomological problem but they are not local.
\newline
We can make another use of the covariant solution \cite{rfm}: $a_2$ is in
general
(before taking the limit $A_2$=$0$) a functional of $\hat{A}^{\dagger}$ and
$\hat{A}$. If we put, formally, $\hat{A}^{\dagger}$=$0$ taking
$\hat{A}$ $\neq$ $0$, $\hat{A}\in SU(N)$, it is clear that we recover the
consistent anomaly. This is not surprising because $\hat{A}^{\dagger}$=$0$
corresponds to $r$=$0$. In some sense the covariant solution of the
extended problem is more general because, using a formal limit, one can
obtain both anomalies. We will explore this property also in the
gravitational case.
\newline
At the end of this section we want to discuss briefly what happens in
presence of a non trivial principal bundle. In this case we cannot define a
determinant smoothly connected to the free one, because there is no
connection on $P(M,G)$ that descends to the null one on the manifold,
hence a
problem of normalization occurs. Moreover the operator $T(r)\DI$ for
$r=0$ is not the correct one: non trivial $P(M,G)$ means that on the
manifold there is no global definition for the connections. One has
different expressions on different patches covering the manifold: these
different expressions are linked by some gauge transformation (transition
functions). Conversely an operator depending on the connections does not
possess a unique form but in different patches has different
representations connected by a gauge transformation.
For example let us suppose that on a subset $\alpha$ $\subset M$ the Dirac
operator is $\DI$ and in $\beta\subset M$ is $\DI\,^{\prime}$ when
$\alpha\cap\beta\neq\circ\hspace{-6pt}/$ being
\[A^{\prime}=U^{-1}AU+U^{-1}dU;\]
it results
\[\DI\,^{\prime}=U^{-1}\DI\, U.\]
Now in order to have a determinant we need the eigenvalues, that are global
properties of the operator: the transformation law means that if in
$\alpha$
\[\alpha : \DI\,\Psi_n=\lambda_n\Psi_n,\]
in $\beta$
\[\beta : \DI\,^{\prime}\Psi^{\prime}_n=\lambda_n\Psi^{\prime}_n\]
with $\Psi^{\prime}_n=U\Psi_n$. In this way the eigenvalues do not depend
on the patches.
\newline
On a non-trivial $P(M,G)$ the Weyl operator does possess the correct
transformation law passing between different patches: but for the
eigenvalue problem the relevant operator is $T\DI$ and we need
\[(T\DI\,)^{\prime}=U^{-1}(T\DI\,)U.\]
This relation forces upon $T$
\[ T^{\prime}=U^{-1}TU.\]
One immediately realizes that $T$ must depend on some fixed background
connection $A_0$, belonging to $P(M,G)$. The natural choice is
\[T=T(0)=(\partiall+A\hspace{-6pt}/_{0}^{\dagger}).\]
Nevertheless we can again define a family of operators $T(r)$ interpolating
between the consistent and the covariant case
\[T(r)=i\tilde{\sigma}_{\mu}(\partial_{\mu}+(1-r)\hat{A}_{0\mu}^{\dagger}+
r\hat{A}_{\mu}^{\dagger}).\]
It is very easy to verify that
\[T^{\prime}(r)= U^{-1}T(r)U.\]
For $r\!=\!1$ the dependence on $A_0$ disappears, so that the covariant anomaly
does not depend on the background fixed connection according to the
cohomological argument of reference [7]
\newline
{\bf Example: $d=2$ non abelian gauge theory}
\newline
The relevant Seeley-de Witt coefficients are:
\[H_{1}(r)=-\frac{1}{2}r[\hat{A}_{\mu},\hat{A}^{\dagger}_{\mu}]+\frac{i}{2}\,r
\epsilon_{\mu\nu}[\hat{A}_{\mu},\hat{A}^{\dagger}_{\nu}],\]
\EQ
-\frac{1}{2}\partial_{\mu}[r\hat{A}^{\dagger}_{\mu}-\hat{A}^{\mu}]+\frac{i}{2}
\epsilon_{\mu\nu}\partial_{\mu}(\hat{A}_{\nu}+r\hat{A}^{\dagger}_{\nu}].
\EN
\[\tilde{H}_{1}(r)=\frac{1}{2}r[\hat{A}_{\mu},\hat{A}^{\dagger}_{\mu}]
+\frac{i}{2}\,r\epsilon_{\mu\nu}[\hat{A}_{\mu},\hat{A}^{\dagger}_{\nu}]\]
\EQ
+\frac{1}{2}\partial_{\mu}[r\hat{A}^{\dagger}_{\mu}-\hat{A}^{\mu}]-\frac{i}{2}
\epsilon_{\mu\nu}\partial_{\mu}(\hat{A}_{\nu}+r\hat{A}^{\dagger}_{\nu}].
\EN
The choice $r=0$ gives:
\EQ
a_{1}(\alpha)=ia_{2}(\alpha)=\int
d^{2}x\frac{i}{4\pi}Tr[(\epsilon_{\mu\nu}\partial_{\mu}\hat{A}_{\nu}-
i\partial_{\mu}\hat{A}_{\mu})\alpha] \label{eq:marta}
\EN
that, for $\hat{A}_{\mu}\rightarrow A_{\mu}$, is the usual consistent
anomaly (modulo a coboundary). For $r=1$:
\[a_{1}(\alpha)=0,\]
\[a_{2}(\alpha)=\frac{1}{4\pi}\int
d^{2}xTr([\epsilon_{\mu\nu}\partial_{\mu}(\hat{A}_{\nu}+\hat{A}^{\dagger}_{\nu})
+\epsilon{\mu\nu}[\hat{A}^{\dagger}_{\mu},\hat{A}_{\nu}]-\]
\EQ
-i\partial_{\mu}(\hat{A}_{\mu}-
\hat{A}^{\dagger}_{\mu})-i[\hat{A}^{\dagger}_{\mu},\hat{A}_{\mu}]]\alpha),
\EN
and, after the real projection, $\hat{A}^{\dagger}_{\mu}=\hat{A}_{\mu}$
\[a_{2}(\alpha)=\frac{1}{4\pi}\int d^{2}x
Tr[(\epsilon_{\mu\nu}F_{\mu\nu})\alpha]=\hat{a}_{cov}.\]
If we put formally in $a_{2}(\al)$
\[ \hat{A}^{\dagger}_{\mu}=0,\]
we recover the consistent anomaly (\ref{eq:marta}) (modulo a factor i).
\vfill\eject

\section{Gravitational anomalies}
In this section we turn our attention to the anomalous behaviour of chiral
spinors coupled to a Riemannian background. The occurrence of gravitational
anomalies has been pointed out by the pioneering work of Alvarez-Gaum\`{e}
and Witten \cite{rfu}, and it has attracted much attention in the middle of the
eighties \cite{rfv1,rfv2,rfv3}:
for more recent studies see \cite{rfz}. We are interested
in generalizing the formalism of covariant anomalies in this context and in
finding an analytical definition of the curved Weyl determinant, using the
properties of the extended algebra. We have not been able to give
a general solution beyond $d=2$ so far: nevertheless we have performed an
explicit
coordinate-invariant calculation of the determinant in the two dimensional
space, pointing out the correct operator to use on a $\zeta$-function approach.
\newline
\subsection{General properties of gravitational anomalies}
The classical Weyl action is
\EQ
S_{cl}=\inta\bar{\psi}\,e^{\mu}_{a}\,\s_{a}i\,(\partial_{\mu}+
\frac{1}{4}\Omega_{\mu cd}\ts_{c}\s_{d})\,\psi \label{eq:grav}
\EN
where $\s^{a}$ and $\ts^{b}$ are the Weyl matrices, $E_{\mu a}$ are the
n-beins fields (with inverse $e^{\mu}_{a}$) and $\Omega_{\mu cd}$ is the
spin-connection linked to the metric tensor
\[ g_{\mu\nu}=E_{\mu a}\,E_{\nu a}\]
\[\Omega_{\mu ab}=e^{\nu}_{a}\,(\partial_{\mu}E_{\nu b}-\Gamma^{\lambda}
_{\mu\nu}E_{\lambda b}).\]
We assume the absence of torsion, so $\Gamma^{\lambda}
_{\mu\nu}$ is the usual Levi-Civita connection.
\newline
We are interested in describing the symmetries of the action (\ref{eq:grav}):
for the moment we do not worry about the global description of this
transformations on which we will make some comments in the sequel. We
remark that, at variance with the gauge case, the symmetries act on the
fibers as well as on the base manifold.
\newline
{\bf Coordinate transformations (diffeomorphisms)}:
\[\delta E_{\mu a}(x)=\partial_{\mu}\xi^{\nu}(x)\,E_{\nu a}(x)+\partial_{\nu}E
_{\mu a}(x)\,\xi^{\mu}(x),\]
\EQ
\delta \psi(x)=\partial_{\mu}\psi(x)\,\xi^{\mu}(x),
\EN
$\xi^{\mu}$ generates the infinitesimal diffeomorphism on the base manifold.
\newline
{\bf Frame rotations ($SO(2n,R)$ transformations):}
\[\de E_{\mu a}(x)=\lambda_{ab}(x)E_{\mu b}(x),\]
\EQ
\de\psi(x)=\lambda_{ab}(x)\omega_{ab} \psi(x),
\EN
with $\lambda_{ab}=-\lambda_{ba}$ and $\omega_{ab}$ the generators of $SO(
2n,R)$ in the chosen spinorial representation (really the relevant group
is $Spin(2n)$ that is a double covering of $SO(2n,R)$).For the Weyl case
\[\omega_{ab}=\frac{1}{4}\,[\s_{a}\ts_{b}-\s_{b}\ts_{a}].\]
{\bf Conformal transformations:}
\[\de E_{\mu a}(x)=\lb (x)E_{\mu a}(x),\]
\EQ
\de\psi(x)=\nu\lambda(x)\psi(x), \label{eq:gigi}
\EN
where $\nu=n\!-\!\frac{1}{2}$.
\newline
Let $\Gamma[E]$ be the vacuum functional of the theory in the external
background: let us study the Ward identity derived from $SO(2n,R)$ symmetry
and diffeomorphism invariance. Classically there are two currents:
\newline
the spin-current
\EQ
J^{\mu}_{ab}(x)=\frac{\de S_{cl}}{\de \Omega_{\mu ab}(x)}
\EN
and the consistent energy-momentum tensor
\EQ
T^{\mu}_{ a}(x)=\frac{\de S_{cl}}{\de E_{\mu a}(x)}.
\EN
{}From $T^{\mu}_{a}$ one can construct the symmetric energy-momentum tensor
\EQ
T^{\mu\nu}=\frac{1}{2}\,(e^{\mu}_{a}T^{\nu}_{a}+e^{\nu}_{a}T^{\mu}_{a})
\EN
and
\EQ
T_{ab}=\frac{1}{2}\,(E_{\mu a}T^{\mu}_{b}-E_{\mu b}T^{\mu}_{a}).
\EN
In absence of torsion
\EQ
T_{ab}=0
\EN
and the classical symmetries give the equations:
\bea
D_{\mu}J^{\mu}_{ab}=0\label{eq:pero},\\
D_{\mu}T^{\mu\nu}=0.
\eea
The presence of anomalies changes these conservation laws: let us suppose
that $\Gamma[E]$ is not invariant under frame rotation. The generator of
the $SO(2n,R)$ rotations is
\[\de(\lambda)=\inta (\de_{\lambda}E_{\mu a})\,\frac{\de}{\de E_{\mu a}},\]
\[\de_{\lambda}E_{\mu a}=-\lambda_{ab}E_{\mu b},\]
giving
\EQ
\de(\lambda)\Gamma[E]=\inta \lambda_{ab}\,G_{ab}=t(\lambda).
\EN
But, using the quantum definition of the consistent energy-momentum tensor
\[\frac{\de\Gamma[E]}{\de E_{\mu a}}=T^{\mu}_{a},\]
\EQ
\de(\lb)\Gamma[E]=\inta \lb_{ab}T_{ab}.
\EN
The anomaly $G_{ab}$ is identified with the antisymmetric part of the
energy-momentum tensor (classically zero): the equation (\ref{eq:pero}) is
modified:
\EQ
D_{\mu}J^{\mu}_{ab}+T_{ab}=0.
\EN
The covariant divergence of the energy-momentum tensor involves the
diffeomorphism invariance: let $\de(\xi)$ be the generator of this symmetry
\[\de(\xi)=\inta (\de_{\xi}E_{\mu a})\frac{\de}{\de E_{\mu a}}=\]
\EQ
=\inta \frac{1}{2}(\de_{\xi}g_{\mu\nu})e^{\nu}_{a}\frac{\de}{\de E_{\mu a}}+
(\de_{\lb(\xi)}E_{\mu a})\frac{\de}{\de E_{\mu a}},
\EN
with
\bea
\de_{\xi}g_{\mu\nu}&\,=\,&D_{\mu}\,\xi_{\nu}+D_{\nu}\,\xi_{\mu},\nonumber\\
\lb_{ab}(\xi)&\,=
\,&\frac{1}{2}e^{\nu}_{a}\,e^{\mu}_{b}(\partial_{\mu}\xi_{\nu}
+\partial_{\nu}\xi_{\mu})-\xi^{\lb}\Omega_{\lb ab}. \label{eq:rossa}
\eea
Diffeomorphism anomaly means
\EQ
\de(\xi)\Gamma[E]=\inta \xi^{\mu}a_{\mu}=a(\xi)
\EN
Using again the definition of $T^{\mu}_{a}$ and assuming the presence of
frame anomaly it is easy to prove
\EQ
D_{\mu}T^{\mu\nu}+l^{\nu}=a^{\nu}, \label{eq:leo}
\EN
where
\EQ
l^{\mu}=-D_{\nu}\,(e^{\nu}_{a}e^{\mu}_{b}T_{ab})+\Omega^{\nu}_{ab}T_{ab}.
\EN
We remark the interplay between the two anomalies: in fact $a^{\mu}$ is the
genuine diffeomorphism anomaly, manifesting the non-invariance of the Weyl
determinant under coordinate transformations. Nevertheless in the Ward
identity for the energy-momentum tensor, also the Lorentz anomaly
$T_{ab}$ appears, thereby changing the balance equation. Even in absence of
diffeomorphism anomaly, say $a^{\mu}=0$, the energy-momentum tensor is not
covariantly conserved, but this is not a signal of general covariance
breaking. It is a consequence of the Lorentz anomaly: the n-beins fields, not
dynamical at the classical level, acquire a dynamical meaning in presence
of the frame anomaly. This mixing is better understood if we write the
whole algebra of the two symmetries.
\newline
If we try to give a global description of this algebra on the base manifold
we have to be careful if the topology is not trivial (namely if the manifold
is not parallelizable). The key point is that the action of $\xi^{\mu}$ on
the tangent plane (on $E_{\mu a}$ and $\Omega_{\mu ab}$) is defined only up
to a local rotation of the orthogonal frame. Following \cite{rfv1} we introduce
from the beginning a fixed background field $E_{\mu a}^0$ so that the
transformation laws globally defined are:
\[ \de(\lb)E_{\mu a}=\lb_{ab}E_{\mu b},\]
\EQ
\de(\lb)\Omega_{\mu ab}=D_{\mu}\lb_{ab},
\EN
\[ \de(\xi)E_{\mu a}=L_{\xi}E_{\mu a}+(\xi^{\lb}\Omega^{0}_{\lb ab})E_{\mu
b},\]
\EQ
\de(\xi)\Omega_{\mu ab}=L_{\xi}\Omega_{\mu ab}+D_{\mu}[\xi^{\nu}(\Omega_
{\nu ab}-\Omega^{0}_{\nu ab})].
\EN
The following commutation rules hold:
\[[\de(\lb_{1}),\de(\lb_{2})]=\de([\lb_{1},\lb_{2}]),\]
\[[\de(\xi_{1}),\de(\lb_{2})]=\de(\xi^{\mu}D^{0}_{\mu}\lb),\]
\EQ
[\de(\xi_{1}),\de(\xi_{2})]=\de([\xi_{1},\xi_{2}]_{L}\,)-\de(\xi_{1}^{\mu}
\xi_{2}^{\nu}R^{0}_{\mu\nu}),
\EN
where $L_{\xi}$ is the usual Lie derivative \[D^{0}_{\mu}=D_{\mu}+[\Omega^{
0}_{\mu}, \;\;]\]
\[[\xi_{1},\xi_{2}]_{L}^{\mu}=
\xi^{\nu}\partial_{\nu}\xi_{2}^{\mu}-\xi_{2}^{\nu}\partial_{\nu}\xi_{1}^{\mu}\]
 and
\[R_{\mu\nu ab}^{0}=\partial_
{\mu}\Omega^{0}_{\nu ab}-\partial_{\nu}\Omega^{0}_{\mu ab}+[\Omega^{0}_{\mu},
\Omega^{0}_{\nu}]_{ab}.\]
\newline
We notice that the global gravitational algebra is much more complicated than
the gauge one: the W-Z consistency conditions are:
\[\de(\lb_{1})t(\lb_{2})-\de(\lb_{2})t(\lb_{1})=t([\lb_{1},\lb_{2}]),\]
\[\de(\xi_{1})t(\lb_{2})-\de(\lb_{2})a(\xi_{1})=t(\xi^{\mu}D^{0}_{\mu}\lb),\]
\EQ
\de(\xi_{1})t(\lb_{2})-\de(\xi_{2})a(\xi_{1})=a([\xi_{1},\xi_{2}]_{L})-
t(\xi_{1}^{\mu}\xi_{2}^{\mu}R^{0}_{\mu\nu}).  \label{eq:stora}
\EN
In the presence of non-trivial topology is not consistent to assume $a=0$,
while it is possible to set $t=0$. In reference \cite{rfv1} a
solution of the full cohomological problem (\ref{eq:stora}) has been given,
using the same method outlined for the gauge theory: this solution, in the
case of parallelizable manifold, gives a vanishing diffeomorphism anomaly.
It is also possible to construct a Wess-Zumino-Witten action [16], using
the n-bein $E_{\mu a}$, in order to compensate the Lorentz anomaly: the
price to pay is anyway the occurrence of a coordinate anomaly. In the following
we make the hypothesis,
unless special remarks, of working with parallelizable manifold,
disregarding problems of globality. Within this limitation it is consistent to
assume either $a^{\mu}=0$ or $T_{ab}=0$ (one can pass from a situation to
another one with a W-Z-W term): because our operatorial approach will be
manifestly coordinate invariant we take $a^{\mu}=0$. The problem reduces to
finding a solution of:
\[\de(\lb_{1})t(\lb_{2})-\de(\lb_{2})t(\lb_{1})=t([\lb_{1},\lb_{2}]),\]
\EQ
\de(\xi)t(\lb)=-t(\xi^{\mu}\partial_{\mu}\lb).
\EN
The first line is equivalent to a problem similar to the one occurring in
gauge theories
(with gauge group $SO(2n,R)$) while the second equation shows
that $t(\lb)$ transforms as a scalar under diffeomorphism: consistent
Lorentz anomaly can be obtained from the general solution (\ref{eq:ras})
where:
\[ A\rightarrow \Omega=\Omega_{\mu ab}S_{ab}dx^{\mu},\]
\[ F\rightarrow  R=d\Omega+[\Omega,\Omega],\]
$S_{ab}$ are the generator of $SO(2n,R)$.
\newline
A strong property can be derived from the theory of the ad-invariant
polynomials over a Lie algebra: the  polynomial relevant to extract the $
SO(2n,R)$ anomaly is one of rank $n+1$ and it does exist only if $n+1=2k$
(it is not difficult to understand this ``selection rule'' in term of
symmetrized trace of the $SO(2n,R)$ generators). The result is the absence
of Lorentz anomaly in $d=4k$.
\subsection{The covariant Lorentz anomaly}
The presence of the frame anomaly, as in the gauge case, changes the
tensorial properties of the spin-current $J_{\mu ab}$ and of the
energy-momentum tensor $T^{\mu\nu}$. Classically $J_{\mu ab}$ transforms
like
\EQ
\de(\lb)J_{\mu ab}=-(\lb_{ac}J_{\mu cb}+\lb_{cb}J_{\mu ca})=\de_{\lb}J_{\mu
ab}.
\EN
It is not difficult to show that in presence of anomaly:
\[\de(\lb)J_{\mu ab}=\de(\lb)\frac{\de}{\de\Omega_{\mu ab}}\Gamma[E]\]
\EQ
=\de_{\lb}J_{\mu ab}+\frac{\de}{\de\Omega_{\mu ab}}\inta\lb_{cd}T_{cd}.
\EN
We see the appearance of an inhomogeneous term; one of course can define a
covariant spin-current
\[\hat{J}_{\mu ab}=J_{\mu ab}+X_{\mu ab},\]
\EQ
\de(\lb)X_{\mu ab}=-\frac{\de}{\de\Omega_{\mu ab}}\inta\lb_{cd}T_{cd},
\EN
modifying the anomaly
\EQ
D_{\mu}\hat{J}^{\mu}_{ab}=-T_{ab}+D_{\mu}X^{\mu}_{ab}\equiv\hat{T}_{ab}.
\EN
The energy-momentum tensor has the same behaviour:
\[T^{\mu\nu}=\frac{1}{2}(e^{\mu}_{a}\frac{\de}{\de E_{\nu a}}+e^{\nu}_{a}
\frac{\de}{\de E_{\mu a}})\Gamma[E],\]
\EQ
\de(\lb)T^{\mu\nu}=
[\de(\lb),\frac{1}{2}(e^{\mu}_{a}\frac{\de}{\de E_{\nu a}}+e^{\nu}_{a}
\frac{\de}{\de E_{\mu a}})]\Gamma[E]+
\frac{1}{2}(e^{\mu}_{a}\frac{\de}{\de E_{\nu a}}+e^{\nu}_{a}
\frac{\de}{\de E_{\mu a}})\de(\lb)\Gamma[E].
\EN
The commutator vanishes and therefore
\EQ
\de(\lb)T^{\mu\nu}=\frac{1}{2}(e^{\mu}_{a}\frac{\de}{\de E_{\nu a}}+e^{\nu}_{a}
\frac{\de}{\de E_{\mu a}})\inta\lb_{cd}T_{cd},
\EN
that is in general different from zero: $T^{\mu\nu}$ is not a tensor. One can
again redefine it, adding a local polynomial
\EQ
\hat{T}^{\mu\nu}=T^{\mu\nu}+l^{\mu\nu}
\EN
and thereby recovering covariance with
\[\de(\lb)l^{\mu\nu}=-\de(\lb)T^{\mu\nu}.\]
The balance equation (\ref{eq:leo}) is changed; we will call the
covariant divergence of $\hat{T}^{\mu\nu}$ covariant anomaly of the
energy-momentum tensor:
\EQ
D_{\mu}\hat{T}^{\mu\nu}=\hat{a}^{\nu}.
\EN
We remark that the presence of the covariant anomaly does not mean a
breaking of the diffeomorphism invariance, as we have seen before.
\newline
To obtain algebrically the Lorentz covariant anomaly $\hat{T}_{ab}$ we
can work in perfect analogy with the gauge case (the covariant
divergence of the energy-momentum tensor actually is not produced by an
algebraic method, but we will compute it directly from a functional
representation): we extend the orthogonal group $SO(2n,R)$ to the complex
orthogonal group $SO(2n,C)$: in order to preserve the relation between
n-bein and metric we choose a particular complexification of the external
fields. We start with a complex n-bein:
\[\hat{E}_{\mu a}=E_{\mu a}^{1}+i\,E_{\mu a}^{2},\]
with
\[\hat{E}_{\mu a}\hat{E}_{\nu a}=g_{\mu\nu}.\]
The simplest way to achieve this is by taking
\[\hat{E}_{\mu a}=\Lambda_{ab}E_{\mu b},\]
\[\Lambda_{ab}\in SO(2n,C).\]
The relation between $\hat{E}_{\mu a}$ and $g_{\mu\nu}$ gives the
constraints :
\[E_{\mu a}^{1}E_{\nu a}^{1}-E_{\mu a}^{2}E_{\nu a}^{2}=g_{\mu\nu},\]
\[e^{1 \mu}_{ a}E_{\nu a}^{1}-e^{2 \mu}_{ a}E_{\nu
a}^{2}=\de^{\mu}_{\nu},\]
\EQ
e^{1 \mu}_{a}E_{\mu b}^{1}-e^{2 \mu}_{a}E_{\mu b}^{2}=\de_{ab},
\EN
and
\[E_{\mu a}^{1}E_{\nu a}^{2}+E_{\nu a}^{1}E_{\mu a}^{2}=0,\]
\[e^{1 \mu}_{a}E_{\nu a}^{2}+e^{2 \mu}_{a}E_{\nu a}^{1}=0,\]
\EQ
e^{1 \mu}_{a}E_{\mu b}^{2}+e^{2 \mu}_{b}E_{\mu a}^{2}=0.
\EN
We derive the spin-connection
\[\hat{\Omega}_{\mu ab}=\hat{e}^{\nu}_{a}(\partial_{\mu}\hat{E}_{\nu b}-
\Gamma^{\lb}_{\mu\nu}\hat{E}_{\lb b}),\]
\[\hat{\Omega}_{\mu ab}=\Omega^{1}_{\mu ab}+i\,\Omega^{2}_{\mu ab},\]
\[\Omega^{1}_{\mu ab}=(e^{1\nu}_{a}\partial_{\mu}E_{\nu b}^{1}-e^{2\nu}_{a}
\partial_{\mu}E_{\nu b}^{2}-\Gamma^{\lb}_{\mu\nu}(e^{1\nu}_{a}
E^{1}_{\lb b}-e^{2 \nu}_{a}E^{1}_{\lb b}),\]
\EQ
\Omega^{2}_{\mu ab}=(e^{1\nu}_{a}\partial_{\mu}E_{\nu b}^{2}+e^{2\nu}_{a}
\partial_{\mu}E_{\nu b}^{1})-\Gamma^{\lb}_{\mu\nu}(e^{1 \nu}_{a}E^{2}_{\lb b}
-e^{2 \nu}_{a}E^{1}_{\lb b}).
\EN
If we call, like in the gauge case, $\de_{1}(\lb)$ the transformation
of the maximal compact subgroup ($SO(2n,R)$) and $\de_{2}(\lb)$ the one of
its orthogonal complement in $SO(2n,C)$, we obtain:
\[ \de_{1}(\lb)e^{1 \mu}_{a}=\lb_{ab}e^{1 \mu}_{b},\]
\[ \de_{1}(\lb)e^{2 \mu}_{a}=\lb_{ab}e^{2 \mu}_{b},\]
\[ \de_{1}(\lb)\Omega^{1}_{\mu ab}=-\partial_{\mu}\lb_{ab}+\lb_{ac}
\Omega^{1}_{\mu cb}+\lb_{bc}\Omega^{1}_{\mu ac},\]
\EQ
\de_{1}(\lb)\Omega^{2}_{\mu ab}=\lb_{ac}\Omega^{2}_{\mu cb}+\lb_{bc}
\Omega^{2}_{\mu ac}.
\EN
\newline
\[ \de_{2}(\lb)e^{1 \mu}_{a}=\lb_{ab}e^{2 \mu}_{b},\]
\[ \de_{2}(\lb)e^{2 \mu}_{a}=-\lb_{ab}e^{1 \mu}_{b},\]
\[ \de_{2}(\lb)\Omega^{1}_{\mu ab}=\lb_{ac}\Omega^{2}_{\mu cb}+\lb_{bc}
\Omega^{2}_{\mu ac},\]
\EQ
\de_{2}(\lb)\Omega^{2}_{\mu ab}=\partial_{\mu}\lb_{ab}-\lb_{ac}
\Omega^{1}_{\mu cb}-\lb_{bc}\Omega^{1}_{\mu ac},
\EN
$\de_{1}$ and $\de_{2}$ generate the $SO(2n,C)$ algebra
\[ [\de_{1}(\lb_{1}),\de_{1}(\lb_{2})]=\de_{1}([\lb_{1},\lb_{2}]),\]
\[ [\de_{2}(\lb_{1}),\de_{2}(\lb_{2})]=-\de_{1}([\lb_{1},\lb_{2}]),\]
\[[\de_{1}(\lb_{1}),\de_{2}(\lb_{2})]=\de_{2}([\lb_{1},\lb_{2}]).\]
The W-Z consistency conditions for $SO(2n,C)$ are similar to the
$SL(n,C)$ ones, studied before:
\[ \de_{1}(\lb)\Gamma=t_{1}(\lb),\]
\[ \de_{2}(\lb)\Gamma=t_{2}(\lb),\]
with the consistent solution:
\[t_{1}(\lb)=\pm it_{2}(\lb)\]
and the covariant solution:
\[ t_{1}(\lb)=0,\]
\[ t_{2}(\lb)\neq0.\]
In the limit $\Omega^{2}_{\mu ab}=0$ we recover the known expression for the
consistent and covariant Lorentz anomaly.
\subsection{The Weyl determinant in curved space}
The relevant object in the study of $\Gamma[E]$ is the operator
appearing in the action (\ref {eq:grav}):
\EQ
\DI=i\s_{a}e^{\mu}_{a}\,(\,\partial_{\mu}+\frac{1}{4}\Omega_{\mu cd}\tilde
{\s}_{c}\s_{d}\,)
\EN
acting on the space $\Gamma(S_{+})$; the invariant measure is $d^{2n}x\sqrt
{g}$. We begin by studying the transformation properties of $\DI$ and $\tilde{
\DI}$.
\newline
{\bf Frame rotation}:
\[\de(\lb)\DI=\DI\,\tilde{\tau}-\tau\DI,\]
\EQ
\de(\lb)\tilde{\DI}=\tilde{\DI}\,\tau-\tilde{\tau}\tilde{\DI}
\label{eq:rollo}
\EN
with
\[\tau=\frac{1}{4}\lb_{ab}\tilde{\s}_{a}\s_{b},\]
\[\tilde{\tau}=\frac{1}{4}\lb_{ab}\s_{a}\tilde{\s}_{b}.\]
{\bf Diffeomorphism:}
\[ \de(\xi)\DI=\xi^{\mu}\partial_{\mu}\DI-\DI\,\xi^{\mu}\partial_{\mu},\]
\EQ
\de(\xi)\tilde{\DI}=\xi^{\mu}\partial_{\mu}\tilde{\DI}-
\tilde{\DI}\,\xi^{\mu}\partial_{\mu}.
\EN
As we have seen before it is not possible to define directly, by $\zeta$-
function regularization, the determinant of $\DI\,$:
in order to have a good eigenvalue
problem, we should find the correct isomorphism. One immediately
realizes that it is not possible to use the same trick of the gauge
theories: from the physical point of view we are tempted to introduce a
free partner for the chiral spinor interacting with gravity
\EQ
T=(i\tilde{\s}_{a}\partial_{a}).  \label{eq:pietro}
\EN
In so doing we inevitably break both symmetries (we are looking for a
diffeomorphism invariant theory): moreover (\ref{eq:pietro}) is not a
fixed isomorphism because a general coordinates transformation changes it
and it does not admit a good geometrical interpretation having no invariant
meaning in curved space. One may think about a generalization of the Weyl
kinetic operator:
\EQ
i\tilde{\s}_{a}e^{\mu}_{a}\partial_{\mu}.
\EN
This a good operator, but it is not free, depending crucially on the n-bein
field. Anyway we can try to fix the isomorphism $T$ satisfying the
properties required to represent a solution of the extended W-Z conditions.
Actually we will study an  operator more general than $T\DI$\,: we define
\EQ
\Gamma[\hat{E};r,s]=
\frac{1}{k}\ln Det[\,\tilde{\DI}_{c}\,(r)\,\DI_{c}\,(s)\,]
\label{eq:pluto}
\EN
with
\[ \tilde{\DI}_{c}(r)=i\tilde{\s}_{a}\hat{e}^{\mu}_{a}\,(\,\partial_{\mu}+
\frac{r}{4}\hat{\Omega}^{\dagger}_{\mu cd}\s_{c}\tilde{\s}_{d}\,),\]
\EQ
\DI_{c}(s)=i\s_{a}\hat{e}^{\mu}_{a}\,(\,\partial_{\mu}+
\frac{s}{4}\hat{\Omega}_{\mu cd}\tilde{\s}_{c}\s_{d}\,),
\EN
where we have used the complexification, deforming both operators with $r,
s\in{\bf R}$; $k$ is a suitable normalization factor. The functional is defined
by $\zeta$-function technique:
\EQ
\Gamma[\hat{E};r,s]=\frac{1}{k}\frac{d}{dt}\,Tr[\,(\tilde{\DI}_{c}(r)
\DI_{c}(s)\,)^{-t}\,]_{t=0}.
\EN
Let us note that $r=s=1$ , $k=2$ does correspond to the modulus case:
\EQ
\frac{1}{2}\frac{d}{dt}\,Tr[\,(\DI\,^{\dagger}_{c}\DI_{c})^{-t}\,]_{t=0}=
\ln det |\DI_{c}|.
\EN
The definition (\ref{eq:pluto}) gives a functional invariant under
diffeomorphism: it is easy to show that
\[\de(\xi)(\,\tilde{\DI}_{c}(r)\DI_{c}(s)\,)
=-[\,(\,\tilde{\DI}_{c}(r)\DI_{c}(s)\,),\xi^{\mu}\partial_{\mu}\,]. \]
Using the properties of the complex powers of elliptic operators
\cite{rfs},
namely the fact they are of trace class
\[ Tr[AB]=Tr[BA],\]
this change does not affect the determinant. For frame rotations the
situation is not trivial: it is a simple exercise to prove from
(\ref{eq:rollo})
\[ \de_{1}(\lb)\DI_{c}(s)=s[\DI_{c}(s)\tilde{\tau}-
\tau\DI_{c}(s)]+(s-1)(\tau-\tilde{\tau})\DI_{c}(s)
+(1-s)A(\hat{\tau}),\]
\EQ
\de_{1}(\lb)\tilde{\DI}_{c}(r)=r[\tilde{\DI}_{c}(r)\tau-
\tilde{\tau}\tilde{\DI}_{c}(r)]+
(1-r)(\tau-\tilde{\tau})\tilde{\DI}_{c}(r)+(1-r)\tilde{A}(\tau)
\EN
and
\[ \de_{2}(\lb)\DI_{c}(s)=is[\DI_{c}(s)\tilde{\tau}-\tau\DI_{c}(s)]
+i(s-1)(\tau-\tilde{\tau})\DI_{c}(s)
+i(1-s)A(\tilde{\tau}),\]
\EQ
\de_{2}(\lb)\tilde{\DI}_{c}(r)=ir[\tilde{\tau}\tilde{\DI}_{c}(r)-
\tilde{\DI}_{c}(r)\tau]
-i(1-r)(\tau-\tilde{\tau})\tilde{\DI}_{c}(r)-i(1-r)\tilde{A}(\tau),
\EN
where
\[A(\tilde{\tau})=([\s_{a},\tilde{\tau}]i\hat{e}^{\mu}_{a}\partial_{\mu}
+i\,\frac{s}{4}\hat{\Omega}_{\mu cd}\hat{e}^{\mu}_{a}[\s_{a}\tilde{\s}_{c}
\s_{d},\tilde{\tau}]),\]
\[\tilde{A}(\tau)=([\tilde{\s}_{a},\tau]i\hat{e}^{\dagger\mu}_{a}\partial_{\mu}
+i\,\frac{r}{4}\hat{\Omega}^{\dagger}_{\mu cd}\hat{e}^{\dagger\mu}_{a}
[\tilde{\s}_{a}\s_{c}\tilde{\s}_{d},\tau]).\]
{}From these equations we can obtain the whole variation of
$\tilde{\DI}_{c}(r)\DI_{c}(s)$: applying $\zeta$-function technique
\[t_{1}(\tau)=\frac{1}{k}\frac{d}{dt}[t\,Tr\{(\tilde{\DI}_{c}(r)\DI_{c}(s))^{-t}
[(s-1)\tilde{\tau}-(r-1)\tau]+(\DI_{c}(s)\tilde{\DI}_{c}(r))^{-t}
[(r-1)\tau-(s-1)\tilde{\tau}]\}+\]
\EQ
+t\,Tr\{[(1-s)\tilde{\DI}_{c}(r)A(\tilde{\tau})+(1-r)\tilde{A}(\tau)\DI_{c}(s)]
(\tilde{\DI}_{c}(r)\DI_{c}(s))^{-1-t}\}_{t=0},
\EN
\[t_{2}(\tau)=\frac{i}{k}\frac{d}{dt}[t\,Tr\{(\tilde{\DI}_{c}(r)\DI_{c}(s))^{-t}
[(s+1)\tilde{\tau}+(r-1)\tau]+(\DI_{c}(s)\tilde{\DI}_{c}(r))^{-t}
[-(r+1)\tau-(s+1)\tilde{\tau}]\}+\]
\EQ
+t\,Tr\{[(1-s)\tilde{\DI}_{c}(r)A(\tilde{\tau})-(1-r)\tilde{A}(\tau)\DI_{c}(s)]
(\tilde{\DI}_{c}(r)\DI_{c}(s))^{-1-t}\}_{t=0}.
\EN
\subsection{Covariant solution}
Putting $r=1$, $s=1$ and $k=2$ we obtain:
\[ t_{1}(\lb)=0\]
\[ t_{2}(\lb)\neq0\]
\EQ
t_{2}(\lb)=\frac{i}{(4\pi)^{n}}\inta Tr[ \tilde{\tau}H_{n}(\DI\sdag_{c}
\DI_{c})-\tau H_{n}(\DI_{c}\DI\sdag_{c})]. \label{eq:tir}
\EN
We recall that the heat-kernel coefficients $H_{n}$ are local in the external
field and their derivatives. In the limit $\Omega^{2}_{\mu ab}=0$ the
covariant Lorentz anomaly is recovered: really until now we have not
computed the trace (\ref{eq:tir}) for general $d=2n$, as in the gauge case
(see eq.(63)), because of the greater
complexity of the heat-kernel coefficients. The
basic reason is that while in the gauge case only the leading contribution
in $\s$ matrices to the n-th Seeley-de Witt coefficient survives inside the
trace, in the gravitational one the subleading term also gives, in
general, a
non vanishing effect. Hence we cannot make a direct comparison with the
cohomological result given in term of invariant polynomials: nevertheless
the explicit calculation in $d=2,4,6$ is consistent with our conjecture.
\newline
Now we show how to obtain from this functional
\EQ
\hat{\Gamma}[E^{1},E^{2}]=\frac{1}{2}\frac{d}{dt}Tr[(\DI\sdag_{c}
\DI_{c})^{-t}]_{t=0},
\EN
the covariant energy-momentum tensor $\hat{T}^{\mu\nu}$ and the covariant
spin-current $\hat{J}_{\mu ab}$. Let us define ($Im$ denotes the imaginary
part)
\EQ
\tilde{T}^{\mu\nu}=-Im(\hat{e}^{\mu}_{a}\frac{\de}{\de\hat{E}_{\nu a}}+
\hat{e}^{\nu}_{a}\frac{\de}{\de\hat{E}_{\mu a}})\hat{\Gamma}[E^{1},E^{2}]
\label{eq:ty}
\EN
where the operator acting on $\hat{\Gamma}[E^{1},E^{2}]$, written in term
of $E^{1}_{\mu a}$ and $E^{2}_{\mu a}$, is
\EQ
\Theta^{\mu\nu}=\frac{1}{2}(e^{1\mu}_{a}\frac{\de}{\de E^{2}_{\nu a}}+
e^{1\nu}_{a}\frac{\de}{\de E^{2}_{\mu a}})-
\frac{1}{2}(e^{2\mu}_{a}\frac{\de}{\de E^{1}_{\nu a}}+
e^{2\nu}_{a}\frac{\de}{\de E^{1}_{\mu a}}).
\EN
We want to prove that:
\EQ
\hat{T}^{\mu\nu}=\tilde{T}^{\mu\nu}|_{e^{2\mu}_{a}=0}
\EN
Exactly we will show that:
\begin{enumerate}
\item $\de(\lb)\hat{T}^{\mu\nu}=0$, $\;\;$ with $\de(\lb)$ generator of
$SO(2n,R)
$ rotation: $\hat{T}^{\mu\nu}$ is really a tensor.
\item $D_{\mu}\hat{T}^{\mu\nu}=\hat{a}^{\nu}$, $\;\;$ with $\hat{a}^{\mu}$ the
covariant anomaly of the energy-momentum tensor.
\end{enumerate}
The point (1) is very simple: we note
\EQ
\de(\lb)=\lim_{e^{2\mu}_{a} \rightarrow 0}\de_{1}(\lb)
\EN
where the compact generator has the expression
\[\de_{1}(\lb)=\inta\lb_{ab}(e^{1\mu}_{b}\frac{\de}{\de E^{2}_{\mu a}}+
e^{1\mu}_{b}\frac{\de}{\de E^{2}_{\mu a}}).\]
{}From (\ref{eq:ty})
\EQ
\de_{1}(\lb)\tilde{T}^{\mu\nu}=[\de_{1}(\lb),\Theta^{\mu\nu}]\hat{\Gamma}
+\Theta^{\mu\nu}\de_{1}(\lb)\hat{\Gamma}.
\EN
Being $\de_{1}(\lb)\hat{\Gamma}=0$ (we recall that the modulus is invariant
under the usual $SO(2n,R)$) and using the explicit result $[\de_{1}(\lb),
\Theta^{\mu\nu}]=0$ we obtain:
\EQ
\de_{1}(\lb)\tilde{T}^{\mu\nu}=0.
\EN
This is true for any $e^{2\mu}_{a}$ and in particular for
$e^{2\mu}_{a}=0$.
\newline
The second claim is much subtler. We compute
\EQ
\inta\xi_{\nu}D_{\mu}\tilde{T}^{\mu\nu},
\EN
$\xi_{\nu}$ being an infinitesimal vector field. This expression can be
rewritten as
\EQ
\inta\frac{1}{2}(D_{\mu}\xi_{\nu}+D_{\nu}\xi_{\mu})
(e^{1\mu}_{a}\frac{\de}{\de E^{2}_{\nu a}}-e^{1\mu}_{a}\frac{\de}
{\de E^{2}_{\nu a}})\hat{\Gamma}. \label{eq:pearl}
\EN
Using the expression for the diffeomorphism generated on $\hat{E}_{\mu a}$
and (\ref{eq:rossa}) one recovers the action of $\de(\xi)$ on the real and
the imaginary part of the n-bein:
\EQ
\de(\xi)E^{1}_{\mu a}=\frac{1}{2}(D_{\mu}\xi^{\alpha}+D^{\alpha}\xi_{\mu})
E^{1}_{\alpha a}+\lb^{1}_{ab}E^{1}_{\mu b}-\lb^{2}E^{2}_{\mu b}-
\xi^{\lb}(\Omega^{1}_{\lb ab}E^{1}_{\mu b}-\Omega^{2}_{\lb ab}E^{2}_{\mu b}
)
\EN
\EQ
\de(\xi)E^{2}_{\mu a}=\frac{1}{2}(D_{\mu}\xi^{\alpha}+D^{\alpha}\xi_{\mu})
E^{2}_{\alpha a}+\lb^{1}_{ab}E^{2}_{\mu b}+\lb^{2}E^{1}_{\mu b}-
\xi^{\lb}(\Omega^{1}_{\lb ab}E^{2}_{\mu b}+\Omega^{2}_{\lb ab}E^{1}_{\mu b}
)
\EN
with
\[\lb^{1}_{ab}=\frac{1}{2}(\partial_{\mu}\xi_{\alpha}-\partial_{\alpha}
\xi_{\mu})[e^{1\alpha}_{a}e^{1\mu}_{b}-e^{2\alpha}_{a}e^{2\mu}_{b}],\]
\[\lb^{2}_{ab}=\frac{1}{2}(\partial_{\mu}\xi_{\alpha}-\partial_{\alpha}
\xi_{\mu})[e^{1\alpha}_{a}e^{2\mu}_{b}+e^{2\alpha}_{a}e^{1\mu}_{b}].\]
The last equations allow us  to write (\ref{eq:pearl}) as:
\[ \inta[\de(\xi)E^{1}_{\mu a}\frac{\de}{\de E^{2}_{\mu a}}-
\de(\xi)E^{2}_{\mu a}\frac{\de}{\de E^{1}_{\mu a}}]\hat{\Gamma}+\]
 \[+\inta[\lb^{1}_{ab}-\xi^{\lb}\Omega^{1}_{\lb ab}]
[E^{2}_{\mu b}\frac{\de}{\de E^{2}_{\mu a}}-
E^{1}_{\mu b}\frac{\de}{\de E^{1}_{\mu a}}]\hat{\Gamma}-\]
\EQ
-\inta[\lb^{2}_{ab}-\xi^{\lb}\Omega^{2}_{\lb ab}]
[E^{2}_{\mu b}\frac{\de}{\de E^{2}_{\mu a}}+
E^{1}_{\mu b}\frac{\de}{\de E^{1}_{\mu a}}]\hat{\Gamma}.
\EN
The third term is zero (it is a compact $SO(2n,C)$ rotation on $\Gamma$),
the second one is related to the covariant anomaly $\hat{T}_{ab}$, giving
\EQ
\inta[\lb^{1}_{ab}-\xi^{\lb}\Omega^{1}_{\lb ab}]
[E^{2}_{\mu b}\frac{\de}{\de E^{2}_{\mu a}}-
E^{1}_{\mu b}\frac{\de}{\de E^{1}_{\mu a}}]\hat{\Gamma}=
[\lb^{1}_{ab}-\xi^{\lb}\Omega^{1}_{\lb ab}]\hat{T}^{ab}. \label{eq:pil}
\EN
The operator acting on $\hat{\Gamma}$ on the first line is:
\EQ
[\de(\xi)E^{1}_{\mu a}\frac{\de}{\de E^{2}_{\mu a}}-
\de(\xi)E^{2}_{\mu a}\frac{\de}{\de E^{1}_{\mu a}}]\hat{E}_{\mu a}=
i\de(\xi)\hat{E}_{\mu a}=\de(i\xi)\hat{E}_{\mu a}.
\EN
It generates the diffeomorphism with an imaginary parameter: the functional
$\hat{\Gamma}$ is not invariant under this transformations at variance with a
real diffeomorphism. Let us study the variation of $\DI_{c}$ and
$\DI\sdag_{c}$ under $\de(i\xi)$: after some efforts we get
\EQ
\de(i\xi)\DI_{c}=\hat{d}_{1}+\hat{d}_{2}
\EN
with
\[\hat{d}_{1}=\frac{1}{4}\{D_{\mu},\de_{\xi}g_{\mu\nu}\s_{\nu}\},\]
\[\hat{d}_{2}=\DI_{c}\,\tilde{\tau}-\tau\DI_{c}+\{\frac{1}{4}g^{\mu}
_{\mu},\DI_{c}\},\]
and the parameter in $\tau$ and $\tilde{\tau}$ is
\[\tilde{\lb}_{ab}=\frac{i}{2}\hat{E}_{\mu a}\hat{E}_{\nu
b}\,(\partial^{\mu}\xi^{\nu}-\partial^{\nu}\xi^{\mu})-i\xi^{\mu}\hat{\Omega}
_{\mu ab}.\]
Conversely
\EQ
\de(i\xi)\DI\sdag_{c}=\tilde{d}_{1}+\tilde{d}_{2}
\EN
with
\[\tilde{d}_{1}=\frac{1}{4}\{D_{\mu},\de_{\xi}g_{\mu\nu}\tilde{\s}_{\nu}\},\]
\[\tilde{d}_{2}=\DI\,^{\dagger}_{c}\,\tau^{\dagger}
-\tilde{\tau}^{\dagger}\DI\sdag_{c}+
\{\frac{1}{4}g^{\mu}_{\mu},\DI\,^{\dagger}_{c}\}.\]
The total variation is:
\[\de(i\xi)(\DI\sdag_{c}\DI_{c})=\DI\sdag_{c}(\tau^{\dagger}-\tau)
\DI_{c}+\DI\sdag_{c}\DI_{c}\tilde{\tau}-
\tilde{\tau}^{\dagger}\DI\sdag_{c}\DI_{c}+\]
\EQ
+\frac{i}{4}[\de_{\xi}g^{\mu}_{\mu},\DI\sdag_{c}\DI_{c}]+\tilde{d}_{1}+
\hat{d}_{1}.
\EN
The dilatation part, proportional to $\de_{\xi}g^{\mu}_{\mu}$, drops out
in the variation of the determinant being a commutator. The rotational
part, proportional to $\tau$ or $\tilde{\tau}^{\dagger}$, gives
contribution only when $Im(\tilde{\lb}_{ab})\neq 0$: it can be
considered as the action of the non
compact sector of
$SO(2n,C)$ on $\hat{\Gamma}$, parameterized by
\[-(\lb^{1}_{ab}-\xi^{\mu}\Omega^{1}_{\mu ab}).\]
Therefore $\de(i\xi)\hat{\Gamma}$ gets a first contribution
\EQ
-\inta(\lb^{1}_{ab}-\xi^{\mu}\Omega^{1}_{\mu ab})\hat{T}_{ab}
\EN
that cancels against the second part of (\ref{eq:pil}). It only
remains to calculate the $\tilde{d}_{1}$, $\hat{d}_{1}$ contribution.
After defining:
\EQ
\Lambda=\frac{i}{4}\{D_{\mu},\xi^{\mu}\}+\frac{1}{4}(\tilde{\xi}\DI_{c}+
\DI\sdag_{c}\xi),
\EN
\EQ
\tilde{\Lambda}=\frac{i}{4}\{D_{\mu},\xi^{\mu}\}+
\frac{1}{4}(\tilde{\xi}\DI_{c}+
\DI\sdag_{c}\xi),
\EN
with
\[\xi=\xi^{\alpha}\s_{\alpha},\]
\[\tilde{\xi}=\xi^{\alpha}\tilde{\s}_{\alpha},\]
we have the useful formula:
\EQ
\tilde{d}_{1}+\hat{d}_{1}=\Lambda^{\dagger}\DI\sdag_{c}\DI_{c}-
\DI\sdag_{c}\DI_{c}\Lambda+\DI\sdag_{c}
(\tilde{\Lambda}-\tilde{\Lambda}^{\dagger})\DI_{c}
\EN
derived from the identities
\[\hat{d}_{1}=\tilde{\Lambda}\DI_{c}-\DI_{c}\Lambda,\]
\[\tilde{d}_{1}=\Lambda^{\dagger}\DI\sdag_{c}-\DI\sdag_{c}
\tilde{\Lambda}^{\dagger}.\]
Again the $\zeta$-function is sensitive only to variations coming from
imaginary terms
\EQ
-\frac{i}{4}\{\{D_{\mu},\xi^{\mu}\},\DI\sdag_{c}\DI_{c}\}
+\frac{i}{2}\DI\sdag_{c}\{D_{\mu},\xi^{\mu}\}\DI_{c}.
\EN
At the end we are left with:
\[\inta\xi_{\nu}D_{\mu}\tilde{T}^{\mu\nu}=
-\frac{i}{4}\frac{d}{ds}(sTr\{[(\DI\sdag_{c}\DI_{c})^{-s}-(\DI_{c}\DI\sdag_{c}
)^{-s}]\{D_{\lb},\xi^{\lb}\}\})_{s=0}\]
Some heat-kernel calculations give (after the elimination of the
parameter $\xi$ and the projection on $E^{2}_{\mu a}$)
\[D_{\mu}T^{\mu\nu}=\frac{1}{4(4\pi)^{n}}Tr[D^{x}_{\lb}H_{n}(x,y)-
D^{y}_{\lb}H_{n}(x,y)]_{x=y}g^{\lb\nu}-\]
\EQ
-\frac{1}{4(4\pi)^{n}}Tr[D^{x}_{\lb}\tilde{H}_{n}(x,y)-
D^{y}_{\lb}\tilde{H}_{n}(x,y)]_{x=y}g^{\lb\nu}.
\EN
This expression coincides with the result of \cite{rfh} and can be solved in
term of invariant polynomials recovering the perturbative calculation of
\cite{rfu}.
\newline
For the spin-current we define
\EQ
\hat{J}^{\mu}_{ab}=\frac{1}{i}
\frac{\de}{\de\Omega^{2}_{\mu
ab}}\hat{\Gamma}|_{\Omega^{2}_{\mu ab}=0}.
\EN
Performing a calculation analogous to the one for the energy-momentum
tensor, we find:
\EQ
\de(\lb)\hat{J}^{\mu}_{ab}=\de_{\lb}\hat{J}^{\mu}_{ab}
\EN
and
\EQ
\inta\lb_{ab}(D_{\mu}\hat{J}^{\mu}_{ab})=-\de_{2}(\lb)\hat{\Gamma}=
-\inta\lb_{ab}\hat{T}_{ab}.
\EN
This completes our analysis of the covariant solution of the extended
algebra: from the modulus of the extended Weyl operator we recover the
covariant energy-momentum tensor and spin-current with the relative anomalies.
\subsection{Consistent solution}
We see that, at the variance with the gauge case, the non local terms
do not a priori disappear for some values of $r$
and $s$, unless $r=1=s$ ( the covariant solution). At present we
are unable to find some $r$ and $s$ giving rise in $d=2n$:
\EQ
 t_{1}(\lb)=\pm it_{2}(\lb).
\EN
Besides we do not know if this solution would be local. What we are able to do
is to solve the problem for $n=1$: the abelian character of the theory makes
the commutators $A$ , $\tilde{A}$ to vanish. The calculation is of interest
by itself because we can write explicitly the determinant: the first
computation \cite{rfw} was based on a particular choice of coordinates; in
\cite{rfi}
Leutwyler showed that it is possible to renormalize the effective
action preserving general covariance. In our approach diffeomorphism
invariance appears naturally and the calculation is performed as the
determinant of a well defined elliptic operator, using the $\zeta$-function
technique. We will discuss the problem for a pure gravitational coupling
and then we will generalize the construction of the determinant in presence of
a
 $U(1)$ gauge connection.
 \vfill\eject

\section{Two dimensional Weyl determinant: pure gravitational coupling}
We assume from the beginning that the base manifold is parallelizable: in
this case no problem arises with fixed background connections and the
operators do possess a global expression. Even in this situation,
as simple as it is, the calculation is not completely
straightforward, as we will see: some subtle points must be clarified
before in order to understand the strategy of the computation.
\newline
Firstly we remark that also in the abelian case ($d=2$) there are two
distinct cohomology classes: one can prove that if $t_{2}(\lb)$ realizes
the covariant solution
\begin{eqnarray}
\de_{1}(\lb)t_{2}(\lb_{2})&=&0,\nonumber\\
t_{1}(\lb)&=&0.
\end{eqnarray}
There is no local functional $P$ in the complexified 2-beins
$\hat{E}_{\mu a}$ and $\hat{E}^{\ast}_{\mu a}$ and their derivatives, scalar
under general coordinates transformations, giving the consistent
solution
\EQ
\hat{t}_{2}(\lb)=\pm i\hat{t}_{1}(\lb) \label{eq:rex}
\EN
with
\bea
\hat{t}_{2}(\lb)&=&t_{2}(\lb)+\de_{2}(\lb)P,\nonumber\\
\hat{t}_{1}(\lb)&=&\de_{1}(\lb)P.
\eea
Then we note that the request $t_{2}(\lb)=\pm i\,t_{1}(\lb)$ concerns
the structure of the coboundaries: adding to $\hat{\Gamma}$ some local term
$P$ we obtain a different representative of the cohomology class:
\bea
t_{1}(\lb)\rightarrow \hat{t}_{1}(\lb)=t_{1}(\lb)+\de_{1}(\lb)P,\nonumber\\
t_{2}(\lb)\rightarrow \hat{t}_{2}(\lb)=t_{2}(\lb)+\de_{2}(\lb)P,
\eea
giving in general $\hat{t}_{1}(\lb)\neq \pm i\hat{t}_{2}(\lb)$. Hence it is not
necessary that $\tilde{\DI}_{c}(r)\DI_{c}(s)$ generates anomalies
strictly obeying (\ref{eq:rex}): we require the consistent constraint up
to local terms in the effective action.
\newline
The last remark is about a simplification that our approach assumes in two
dimensions: in order to exploit the solutions of the $SO(2n,C)$ anomaly
algebra we had to work with the full complexified Weyl operator.
The case $d=2$ is very particular and we do not need to make the
complexification. Basically we recognize that
\[Lie\,SO(2,C)\simeq Lie\,(SO(2,R)\otimes R_{+})\]
and we understand $R_{+}$ as the group of conformal transformations,
described in (\ref{eq:gigi}): here conformal means a local dilatation
of the orthogonal frame (we do not touch the base manifold). Actually the
effect of a $SO(2,C)$ rotation on the Weyl operator is the same of a
real rotation plus a conformal transformation.
\newline
In the following we shall use the Weyl matrices
$\s_{1}$=$\tilde{\s}_{1}$=$1$ , $\s_{2}$=$-\tilde{\s}_{2}=i$ and the Weyl
operator
\EQ
\DI=i\s_{a}e^{\mu}_{a}(\partial_{\mu}+i\Omega_{\mu})
\EN
with
\[i\Omega_{\mu}=\frac{1}{4}\Omega_{\mu ab}\tilde{\s}_{a}\s_{b}.\]
An $SO(2,C)$ matrix $\Lambda$ admits the factorization
\[ \Lambda=R\,\hat{\Lambda} \]
where $R\in SO(2,R)$ and
\EQ
\hat{\Lambda}=\left( \begin{array}{cc} \cosh\phi &  -i\sinh\phi \\
                                      i\sinh\phi &  \cosh\phi  \end{array}
                                                              \right)
\EN
from which one can derive the identities
\EQ
(\hat{\Lambda}_{ab}e^{\mu}_{b})\s_{a}=\exp(-\phi)\,e^{\mu}_{a}\s_{a},
\EN
\EQ
(\hat{\Lambda}_{ac}\hat{\Lambda}_{bd}\frac{1}{4}\Omega_{\mu
cd})\tilde{\s}_{a}\s_{b}=i\Omega_{\mu}-\frac{1}{2}\partial_{\mu}\phi,
\EN
\EQ
e^{\mu}_{a}\s_{a}i\partial_{\mu}\phi=e^{\mu}_{a}\s_{a}(\frac{\epsilon^{\mu\nu}}
{\sqrt{g}}\partial_{\nu}\phi).
\EN
The effect of $\hat{\Lambda}$ on $\DI$ is
\[ \hat{\Lambda}:\DI\rightarrow\DI\,'=i\s_{a}e'^{\mu}_{a}(\partial_{\mu}
+i\Omega_{\mu}')\]
with
\EQ
e'^{\mu}_{a}=\exp(-\phi)e^{\mu}_{a},
\EN
\EQ
\Omega'_{\mu}=\Omega_{\mu}+\frac{1}{2}\frac{\epsilon_{\mu}^{\nu}}{\sqrt{g}}
\partial_{\nu}\phi.
\EN
These are the changes of $e^{\mu}_{a}$ and $\Omega_{\mu}$ under a conformal
transformation: the non-compact sector of $SO(2,C)$, represented by
$\hat{\Lambda}$, acts on the Weyl operator as a conformal transformation.
So we do not need to extend the 2-beins to complex values but we will
simply understand $\de(\lb)$ as the conformal generator: in order to match
the parameter $i\tau=\frac{i}{4}\lb_{ab}\tilde{\s}_{a}\s_{b}$ of the
non-compact transformation with the parameter ${\phi}$ we define
\EQ
i\tau=\frac{1}{2}\phi.
\EN
Now the strategy is: after the definition
\EQ
\Gamma[E;r,s]=\frac{1}{k-2}\frac{d}{dt}[Tr(\tilde{\DI}\,(r)\DI\,(s))^{-t}]_{t=0}
\label{eq:fufi}
\EN
with
\[\DI\,(s)=i\s_{a}e^{\mu}_{a}[\partial_{\mu}+\frac{1}{4}(1-s)\Omega_{\mu}],\]
\[\tilde{\DI}\,(r)=i\tilde{\s}_{a}e^{\mu}_{a}[\partial_{\mu}
-\frac{1}{4}(1-r)\Omega_{\mu}],\]
and
\bea
&t_{1}(\lb;r,s)=\de_{1}(\lb)\Gamma[E;r,s],\nonumber\\
&t_{2}(\lb;r,s)=\de_{2}(\lb)\Gamma[E;r,s],
\eea
we want:
\begin{enumerate}
\item To find the values of $r,s,k$ that satisfy the consistently
constraint (up to coboundary terms)
\[t_{1}(\lb;r,s)=\pm it_{2}(\lb;r,s).\]
\item To fix the normalization using the relation between consistent and
covariant anomaly: we obtain the last one acting with $\de_{2}(\lb)$ on
$\ln\sqrt{det(\DI\sdag\DI)}$.
\end{enumerate}
Really we will do something more: in $d=2$ we can compute
$\Gamma[E;r,s]$ for any $r$ and $s$: so we will normalize the
determinant itself using the knowledge of $\ln\sqrt{det(\DI\sdag\DI)}$
($r$=$s$=$0$=$k$). Let us note that (\ref{eq:fufi}) is invariant under
diffeomorphisms, as we have seen, while  the origin of the $SO(2,R)$ breaking
is clear: $\tilde{\DI}\,(r)\DI\,(s)$ does not change
covariantly under rotation of the local frame (in the $\zeta$-function
language the infinitesimal variation of $\tilde{\DI}\,(r)\DI\,(s)$ is
not a commutator). Writing the transformation law
($\tilde{\tau}=-\tau$):
\[\de_{1}(\lb)(\tilde{\DI}\,(r)\DI\,(s))=(r\tau-\tilde{\tau})
(\tilde{\DI}\,(r)\DI\,(s))+\]
\EQ
+(1-s)\tau(\tilde{\DI}\,(r)\DI\,(s))
+\tilde{\DI}\,(r)(-r\tau+s\tilde{\tau})\DI\,(s),
\EN
\newline
\[\de_{2}(\lb)(\tilde{\DI}\,(r)\DI\,(s))=-i[r\tau-\tilde{\tau}]
(\tilde{\DI}\,(r)\DI\,(s))+\]
\EQ
+(\tilde{\DI}\,(r)\DI\,(s))(1-s)\,i\,\tilde{\tau}
+\tilde{\DI}\,(r)i\,[(r-2)\tau+s\tilde{\tau}]\DI\,(s),
\EN
the infinitesimal variation of $\Gamma[E:r,s]$ is:
\[\de_{1}(\lb)\Gamma[E:r,s]=\frac{1}{2-k}Tr[(\tilde{\DI}\,(r)\DI\,(s))^{-t}-
(\DI\,(s)\tilde{\DI}\,(r))^{-t}]_{t=0}(r+s)\tau,\]
\[\de_{2}(\lb)\Gamma[E:r,s]=\frac{i}{2-k}Tr[(\tilde{\DI}\,(r)\DI\,(s))^{-t}
(2+r-s)+
(\DI\,(s)\tilde{\DI}\,(r))^{-t}(2+s-r)]_{t=0}\tau.\]
Using the heat-kernel representation of the complex powers of elliptic
operators \cite{rfs} we obtain
\EQ
\de_{1}(\lb)\Gamma=\frac{1}{2-k}\frac{1}{4\pi}\int d^{2}x\sqrt{g}\,
Tr[H_{1}-\tilde{H}_{1}](r+s)\tau
\EN
\EQ
\de_{2}(\lb)\Gamma=\frac{i}{2-k}\frac{1}{4\pi}\int d^{2}x\sqrt{g}\,
Tr[\,(H_{1}-\tilde{H}_{1})(r-s)+4(H_{1}+\tilde{H}_{1})\,]\tau
\EN
with $H_{1}=H_{1}(\tilde{\DI}\,(r)\DI\,(s))$ and $\tilde{H}_{1}=
H_{1}(\DI\,(s)\tilde{\DI}\,(r))$
$H_{n}$ being the n-th coefficient of the expansion of the heat-kernel in
the limit $t\rightarrow 0$ \cite{rfs}.
\newline
Generalizing the technique developed in [8] to our case (a deformation of
$\DI\sdag\DI \,$) we compute the coefficients $H_{1}$ and $\tilde{H}_{1}$
(we do not report the long but straightforward procedure):
\EQ
H_{1}=\frac{1}{6}R-\frac{1}{8}R(2-s+r)+\frac{i}{2}D_{\mu}\Omega^{\mu}(s+r),
\EN
\EQ
\tilde{H}_{1}=\frac{1}{6}R-\frac{1}{8}R(2+s-r)-\frac{i}{2}D_{\mu}\Omega^{\mu}
(s+r).
\EN
R is the curvature scalar that in $d=2$ is easily expressed as
\EQ
\frac{1}{4}\epsilon_{\mu\nu}\sqrt{g}R=\partial_{\mu}\Omega_{\nu}-
\partial_{\nu}\Omega_{\mu}.
\EN
Then we put:
\[(s-r)=\ua \]
\[(s+r)=\da \]
With this definition and using the explicit form of $H_{1}$ and
$\tilde{H}_{1}$, the ``anomalies'' appear:
\EQ
t_{1}(\lb;r,s)=\frac{1}{4\pi}\frac{1}{2-k}\int d^{2}x\sqrt{g}\,
\tau(\frac{1}{4}\ua \da R+i\, \da^{2} D_{\mu}\Omega^{\mu})
\label{eq:lou}
\EN
\EQ
t_{2}(\lb;r,s)=\frac{i}{4\pi}\frac{1}{2-k}\int d^{2}x\sqrt{g}\,
\tau(\frac{1}{3}R-\frac{1}{4}\ua^{2}R-i\,\ua \da D_{\mu}\Omega^{\mu})
\label{eq:reed}
\EN
Let us look for the covariant solution.
\subsection{Covariant solution}
The condition
\EQ
t_{1}(\lb;r,s)=0
\EN
implies $\da=0$; on $r$ and $s$ we have the constraint:
\EQ
r+s=0.
\EN
The $\de_{2}$ variation is:
\EQ
t_{2}(\lb;r,s)=\frac{i}{4\pi}\frac{1}{2-k}
\int d^{2}x \sqrt{g}\,R\,(\frac{1}{3}-\frac{1}{4}\ua^{2})\,\tau. \label{eq:tip}
\EN
We note that there is no unique solution: varying $r$ and $s$ along
$r$=$-s$ we obtain a continuous family of operators whose determinant
represents
covariant solutions of the extended cohomological problem: they differ
by a coefficient so that we can choose $k$ as a function of $\ua$ in order to
fix the correct normalization. The usual expression is recovered for
the modulus of the Weyl determinant, characterized by $r$=$s$=$k$=$0$
\EQ
t_{2}(\lb)=\frac{i}{24\pi}\int d^{2}x\sqrt{g}\,R\,\tau. \label{eq:tap}
\EN
The normalization condition can be settled from (\ref{eq:tip}) and
(\ref{eq:tap})
\EQ
\frac{1}{4(2-k)}(\frac{1}{3}-\frac{1}{4}\ua^{2})=\frac{1}{24}.
\EN
Solving along $r=-s$ we find:
\[s(k)=\pm\sqrt{\frac{k}{6}},\]
\EQ
r(k)=\mp\sqrt{\frac{k}{6}},
\EN
where $k\geq0$: the $\de_{2}$ variation of the functional (\ref{eq:fufi})
is constant along $[s(k),r(k),k]$. The symmetry
$r,s\rightarrow -r,-s$ reflects a change of $\s$ representation. As we
will see afterwards not only the covariant anomaly but also the functional
itself is constant along $s(k),r(k)$. In other words
\EQ
Det(\tilde{\DI}\,(r(k))\DI\,(s(k)))=[Det(\DI\sdag\DI)]^{2-k}.
\EN
We have found a continuous deformation of the modulus of Weyl determinant,
associated to a continuous family of operators, representing continuous
powers of the modulus itself (the determinants are normalized to the
free laplacian one). The same feature will also appear in the
consistent case. By the way we note that the limit $k=2$ is not
singular: with our normalization we find:
\[Det[\tilde{\DI}\,(\pm\frac{\sqrt{3}}{3})\DI\,(\mp\frac{\sqrt{3}}{3})]=1\]
\subsection{Consistent solution}
The equation (\ref{eq:rex}) gives the system
\[ \frac{1}{4}\ua \da =\pm\,(\frac{1}{3}-\frac{1}{4}\ua^{2})\]
\EQ
y^{2}=\mp \, \ua \da.
\EN
Unfortunately there is no meaningful solution: the only one is
\[\da=0\,;\,\ua^{2}=\frac{4}{3}\]
reducing $t_{1}(\lb)$, $t_{2}(\lb)$ to coboundary terms. We have to exploit
the freedom of inserting a local term in the definition of
$\Gamma[E;r,s]$ to recover the consistent solution: in so doing we
introduce a new parameter, that will appear only in the intermediate
calculation. We add to $\Gamma$
\EQ
P(\al)=\frac{1}{4\pi(2-k)}\al\int d^{2}x\sqrt{g}\, \Omega_{\mu}\Omega^{\mu}
\EN
with $\al\in R$ and we define
\[\Gamma'=\Gamma+P(\al)\]
\[\de_{1}(\lb)\Gamma'=\hat{t}_{1}(\lb)\]
\[\de_{2}(\lb)\Gamma'=\hat{t}_{2}(\lb).\]
It is easy to verify that $P(\al)$ changes (remembering that $\de_{2}$
acts like a conformal transformation):
\EQ
\de_{1}(\lb)P(\al)=\frac{1}{4\pi(2-k)}\al\int d^{2}x\sqrt{g}
\,2\,iD_{\mu}\Omega^{\mu}\tau,
\EN
\EQ
\de_{2}(\lb)P(\al)=\frac{i}{4\pi(2-k)}\al\int d^{2}x\sqrt{g}\,
(-\frac{1}{2}R)\,\tau.
\EN
The new system is:
\[\frac{1}{4}\ua \da=\pm\,(\frac{1}{3}-\frac{1}{4}\ua^{2}+\frac{1}{2}\al)\]
\EQ
\da^{2}+2\al=\mp \, \ua \da.
\label{eq:cicci}
\EN
It leads to the relation
\EQ
\frac{1}{2}\ua \da =\pm\,(\frac{1}{3}-\frac{1}{4}\ua^{2}-\frac{1}{4}\da^{2})
\EN
that will be fundamental and does not depend on $\al$. Solving for
$\ua$ and $\da$ as functions of $\al$ we get:
\[ \da^{2}=3\al^{2},\]
\EQ
\ua^{2}=\frac{4}{3}+3\al^{2}+4\al. \label{eq:birra}
\EN
To fix $\al$ we use the general form of $\Gamma[E;r,s]$. The
calculation is performed in isothermal coordinates \cite{rfhi}: we are allowed
to choose a particular coordinate system, our definition of
determinant being invariant under diffeomorphism. Locally any two dimensional
Riemannian manifold admits a coordinate system in which the metric
tensor has the form
\EQ
g_{\mu\nu}=\exp(4G) \de_{\mu\nu},
\EN
giving the zwei-bein
\EQ
e^{\mu}_{a}=\exp(-2G) \de^{\mu}_{b}(\de_{ab}\cos 2F-\epsilon_{ab}\sin
2F),
\EN
where $F(x)$ describes the freedom of a local orthogonal rotation. We
make the additional assumption of working on a manifold admitting a global
system for this coordinate. In this system the spin-connection and the
scalar curvature have the simple expression:
\[\Omega_{\mu}=\partial_{\mu}F+\epsilon_{\mu\nu}\partial_{\nu}G,\]
\[R=-4\frac{1}{\sqrt{g}}\partial_{\mu}\partial_{\mu}G.\]
Now the possibility to calculate exactly the determinant relies on the
fact that one can write:
\[\tilde{\DI}\,(r)\DI\,(s)=\exp [-G(3-r)+iF(r-1)] (i\partial_{+})
\exp[-G(2+r-s)\]
\EQ
\exp[-iF(r+s)] (i\partial_{-})\exp[G(1-s)-iF(1-s)],
\EN
where $\partial_{\pm}=\partial_{1}\pm\partial_{2}$.
\newline
It is not difficult to find the infinitesimal variation of the
determinant for the transformation
\[G\rightarrow G-\epsilon G,\]
\[F\rightarrow F-\epsilon F,\]
$\epsilon\rightarrow 0$ and to iterate this change driving $G$ and $F$
to zero:
\EQ
Det(\tilde{\DI}\,(r)\DI\,(s))=\exp(-\Gamma[r,s])det(-\partial^{2}).
\label{eq:fre}
\EN
This is the standard decoupling technique \cite{rff} that allows the
calculation
of two dimensional determinants: we note that the normalization to the
free laplacian is a natural bonus of the procedure. Some care is needed
in the iteration for the presence of a conformal factor in the measure.
We give the result deferring the details to the Appendix C. In
isothermal coordinates:
\[
\Gamma'[r,s;\al]=\frac{1}{8\pi(2-k)}\int
d^{2}x\,(\frac{1}{3}-\frac{1}{4}\ua^{2})
\,(4G\partial_{\mu}\partial_{\mu}G)
+i\,(\frac{1}{2}\ua \d)\,(4F\partial_{\mu}\partial_{\mu}G)+\]
\EQ
+\frac{1}{8\pi(2-k)}\int d^{2}x\,(4F\partial_{\mu}\partial_{\mu}F)-
\frac{1}{8\pi(2-k)}\al\int d^{2}x\,[4F\partial_{\mu}\partial_{\mu}F
+4G\partial_{\mu}\partial_{\mu}G].
\EN
We recognize the contribution of $P(\al)$ and $\Gamma[r,s]$. The real
part of this action is not gauge invariant, but we can reduce it to a
gauge invariant form adding a suitable local term, obtaining:
\EQ
\tilde{\Gamma}[r,s;\al]=\frac{1}{8\pi(2-k)}\int d^{2}x\,
(\frac{1}{3}-\frac{1}{4}\ua^{2}-\frac{1}{4}\da^{2})\,(4G\partial_{\mu}
\partial_{\mu}G)
+i\,(\frac{1}{2}\ua \da)\,(4F\partial_{\mu}\partial_{\mu}G)
\EN
\EQ
\tilde{\Gamma}=\Gamma+\frac{1}{8\pi(2-k)}\int d^{2}x\,(\frac{1}{4}\da^{2})
 \,(4F\partial_{\mu}\partial_{\mu}F+4G\partial_{\mu}\partial_{\mu}G).
\EN
In fact it is a general property that the modulus of the Weyl determinant
can always be written as a gauge invariant quantity [19]. Now we can
compare the real part of $\tilde{\Gamma}$ with:
\EQ
\frac{1}{2}\ln Det(\DI\sdag\DI)=\Gamma[r,s]_{r=s=k=0}. \label{eq:jr}
\EN
In other words we require that the definition
\[det(\DI\,)=\exp(-\tilde{\Gamma}[r,s])\]
be consistent with the well known relation
\[|Det(\DI)|=\sqrt{Det(\DI\sdag\DI)}.\]
The solution of the system (\ref{eq:cicci}) implies
\EQ
\tilde{\Gamma}=\frac{1}{8\pi(2-k)}(\frac{1}{2}\ua \da)
\int d^{2}x[4G\partial_{\mu}
\partial_{\mu}G\pm i\,4G\partial_{\mu}\partial_{\mu}F],
\EN
fixing the relative value between the real and the imaginary part of
$\tilde{\Gamma}$ for any $\al$: $\al$ appears only as a device to find
the correct normalization. We remark that $\tilde{\Gamma}$ corresponds
to the ``minimal form'' of the Lorentz anomaly, in which all possible
coboundary terms are subtracted: in this case we cannot have any explicit $\al$
dependence.
Expressing $\frac{1}{2}\ua \da$ as a function of
$\al$ and requiring (\ref{eq:jr})
\EQ
Re\,\tilde{\Gamma}=\Gamma[r,s]|_{r=s=k=0},
\EN
we obtain the equation:
\EQ
\frac{1}{2(2-k)}\ua
\da =\frac{1}{2-k}(-\al-\frac{3}{2}\al^{2})=\frac{1}{6},
\label{eq:vino}
\EN
solved by \[\al=-\frac{1}{3}\pm\frac{1}{3}\sqrt{k-1}\; ; \;k\geq1.\]
\newline
Then, using the definition of $G$ and $F$ and the coordinate invariance
of the determinant, we can write $\tilde{\Gamma}$ in a manifestly
invariant form
\EQ
\tilde{\Gamma}=\frac{1}{192\pi}\int d^{2}x\sqrt{g(x)}\int d^{2}y\sqrt{g(y)}
R(x)\Delta^{-1}_{g}(x,y)R(y)\pm
iR(x)\Delta^{-1}_{g}(x,y)\frac{1}{\sqrt{g}}\partial_{\nu}(\sqrt{g}\Omega^{\nu}
(y))
\EN
with $\Delta^{-1}_{g}(x,y)$ the kernel of the inverse of
Beltrami-Laplace operator:
\EQ
\Delta_{g}=\frac{1}{\sqrt{g}}\partial_{\mu}(g^{\mu\nu}\sqrt{g})\partial_{\nu}.
\EN
$\tilde{\Gamma}$ is connected to $\Gamma$, obtained by a
$\zeta$-function definition, by a local term
\EQ
\beta(k)\int d^{2}x\sqrt{g}\,
\Omega_{\mu}\Omega^{\mu}=\da^{2}\frac{1}{8\pi(2-k)}\int d^{2}x\sqrt{g}\,
\Omega_{\mu}\Omega^{\mu}:
\EN
hence $\tilde{\Gamma}$ is a sort of minimal form of the Weyl effective
action, in which the real part is reduced to the gauge invariant form.
By the way $\tilde{\Gamma}$ coincides with the result of [8], the sign of
the imaginary part being related to the chirality. Coming back to
$\Gamma[r,s]$, expressing through (\ref{eq:birra}) and (\ref{eq:vino})
$r$ and $s$ as function of
$k$, we find a family of $\Gamma[E,k]$, generated by the operators
$\tilde{\DI}\,(r(k))\DI\,(s(k))$, that differ for the allowed local term
and define the Weyl determinant
\EQ
\Gamma[E;r(k),s(k)]=\tilde{\Gamma}+\beta (k) \int
d^{2}x\sqrt{g}\Omega^{\mu}\Omega_{\mu}
\EN
\[\beta(k)=\frac{1}{24\pi}\frac{k-2\sqrt{k-1}}{k-2}\,\,\, ; \,\,\,k\geq 1.\]
We have expressed $\da^{2}$ through $k$: in order to have a well defined
$\Gamma$ for any $k\geq 0$ we have chosen
\[\al=-\frac{1}{3}+\frac{1}{3}\sqrt{k-1}\]
disregarding the other solution. The coefficient of the local term has a
regular behaviour in the limit $k\rightarrow 2$ (it vanishes),
and $\Gamma$ coincides with
$\tilde{\Gamma}$. Solving for $r(k)$ and $s(k)$ we get:
\EQ
r(k)=-\frac{\sqrt{3}}{3}\;\;\; ; \;\;\;
s(k)=\frac{\sqrt{3}}{3}\sqrt{k-1},
\label{eq:zaza}
\EN
\EQ
r(k)=-\frac{\sqrt{3}}{3}\sqrt{k-1}\;\;\;  ;
\;\;\;s(k)=\frac{\sqrt{3}}{3},
\label{eq:zuzu}
\EN
which exhibits a kind of symmetry between the inequivalent
representations of
the Weyl algebra. Again we find a family of operators whose determinants
give powers of the Weyl determinant, up to a local term,
\EQ
Det(\tilde{\DI}\,(r(k))\DI\,(s(k)))=(Det\DI)^{k-2}\exp[-\gamma(k)\int
d^{2}x\sqrt{g}\,\Omega^{\mu}\Omega_{\mu}]
\EN
with $\gamma(k)=\frac{1}{24\pi}(k-2\sqrt{k-1})$ and for $Det\DI$ we mean the
minimal form:
\[Det\DI=exp[-\tilde\Gamma].\]
\newline
In particular one can choose $k=1$ ($s=0$) recovering the original
definition
\[Det\DI=Det(T\DI),\]
\EQ
T=\tilde{\DI}\,(-\frac{\sqrt{3}}{3}).
\EN
We note that for $k=2$ (everything is normalized to the free laplacian)
\EQ
Det(\tilde{\DI}\,(r(2))\DI\,(s(2)))=1.
\EN
It seems that for the critical values $r=\mp \frac{\sqrt{3}}{3}$ ;
$s=\pm\frac{\sqrt{3}}{3}$ the dependence of the determinant on the
external field disappears. At the moment we do not know the
deep reason for this behaviour, if any.
\newline
In this formalism diffeomorphism invariance is manifest; we do not need
any counterterm to recover the general covariance. We can add however a
Wess-Zumino term to $\Gamma$, using the zwei-bein field $E_{\mu
a}$ in order to cancel the Lorentz anomaly, but generating a coordinate
anomaly: we do not know if it is possible to implement it in our
operatorial approach. Another direction of work is to extend the
computation to non parallelizable manifolds.
\newline
At the end we want to show how it is possible to recover the determinant in a
less rigorous but more direct way,
utilizing the formal limit on the complexified connection described
in chapter two. In the gauge case, if we start with
the complexified action, the Weyl determinant is obtained by the
calculation of its modulus, putting formally $\hat{A}_{\mu}^{\dagger}$=$0$
\[ Det(\tilde{\DI}\sdag_{c}\DI_{c})\rightarrow
Det(\tilde{\partiall}\DI).\]
In the gravitational case the same trick does not seem to produce the same
effect: the presence of a n-bein field does not allow to perform a limit on
the spin-connection without altering the n-bein itself, unless we break their
canonical relation. To put $\hat{\Omega}_{\mu ab}^{\dagger}$=$0$ seems to
force $\hat{E}^{\ast}_{\mu a}$=$cost$ while $\hat{E}_{\mu a}\neq 0$: the
general covariance is lost. But in $d$=$2$ the operation can be made in a
particular way to preserve the geometry of the theory. If we complexify the
zwei-bein with a non-compact $SO(2,C)$ rotation
\EQ
\hat{\Lambda}=\left( \begin{array}{cc} \cosh\phi &  -i\sinh\phi \\
                                      i\sinh\phi &  \cosh\phi  \end{array}
                                                              \right)
\EN
the relevant spin-connections are
\[\hat{\Omega}_{\mu}=\Omega_{\mu}+2i\partial_{\mu}\phi,\]
\[\hat{\Omega}^{\dagger}_{\mu}=\Omega_{\mu}-2i\partial_{\mu}\phi,\]
giving
\EQ
\partial_{\mu}\phi=\frac{1}{4i}(\hat{\Omega}_{\mu}-\hat{\Omega}^{\dagger}_{\mu}
)=I_{\mu}. \label{eq:elio}
\EN
As we have seen in the Weyl operator
\[\hat{E}_{\mu a}=\exp(\phi)E_{\mu a},\]
\[\hat{\Omega}_{\mu}=\Omega_{\mu}+\frac{1}{\sqrt{g}}2\varepsilon_{\mu}^{\nu
}\partial_{\nu}\phi,\]
so that
$\hat{E}_{\mu a}$ and $\hat{\Omega}_{\mu}$ are the conformal transformed
objects: here $\phi$ appears as a ``dilaton'' field. In this way $\DI\sdag_{
c}\DI_{c}$ becomes the usual $\DI\sdag\DI$ where the geometric fields are the
conformal transformed of the original ones. Using the explicit result:
\EQ
\frac{1}{2}\ln Det \DI^{\prime}\sdag\DI^{\prime}=\frac{1}{192\pi}\int d^{2}x
\sqrt{g'(x)}\int d^{2}y\sqrt{g'(y)}R'(x)\Delta^{-1}_{g'}(x,y)R'(y)
\EN
with
\[g'_{\mu\nu}=\exp(2\phi)g_{\mu\nu},\]
\[R'=\exp(-2\phi)[R+\Delta_{g}2\phi],\]
and $\Delta_{g'}$=$\Delta_{g}$
we get:
\[\frac{1}{2}\ln Det\DI'\sdag\DI'=\int d^{2}x\sqrt{g(x)}\int d^{2}y\sqrt{g(y)}
[R(x)+2\Delta_{g}\phi]\Delta^{-1}_{g}(x,y)[R(y)+2\Delta_{g}\phi]=\]
\EQ
=\int d^{2}x\sqrt{g(x)}\int d^{2}y\sqrt{g(y)}
R(x)\Delta^{-1}_{g}(x,y)R(y)+\frac{1}{48\pi}\int d^{2}x\sqrt{g}\,[R\phi+
\phi\Delta_{g}\phi].
\EN
At this point we come back to $\hat{\Omega}_{\mu}$ and $\hat{\Omega}_{\mu}^
{\dagger}$: from (\ref{eq:elio}) we have
\EQ
\Delta_{g}\phi=-\frac{1}{\sqrt{g}}\partial_{\mu}(\sqrt{g}I^{\mu})
\EN
giving
\[\frac{1}{2}\ln Det\DI\sdag_{c}\DI_{c}=\frac{1}{48\pi}\int d^{2}x\sqrt{g}\
,I_{\mu}I^{\mu}+\]
\EQ
+\frac{1}{192\pi}\int d^{2}x\sqrt{g(x)}
\int d^{2}y\sqrt{g(y)}R(x)\Delta^{-1}_{g}(x,y)R(y)-
4iR(x)\Delta^{-1}_{g}(x,y)\frac{1}{\sqrt{g}}\partial_{\nu}(\sqrt{g}I^{\nu}
(y)).
\EN
Now all the dependence on the parameter $\phi$ of the complex
transformation is carried by the connections. It easy to check that putting
$\hat{\Omega}_{\mu}$=$0$ or $\hat{\Omega}_{\mu}^{\dagger}$=$0$ one recovers
the correct determinant (with a different sign on the imaginary part).
Really we are left with a local term
\EQ
\frac{1}{768\pi}\int d^{2}x \sqrt{g}\,\Omega_{\mu}\Omega^{\mu}
\EN
that, as usual, can be removed. At the moment we do not know if this
procedure gives the correct result in more than two dimensions:
beyond $d$=$2$ it is not
possible to represent complex rotation as conformal transformations,
but in principle, one can work with the Seeley-de Witt coefficients.
\vfill\eject
\section{The insertion of U(1) gauge potential}
In this section we generalize the construction we have just described
in the presence
of an external electromagnetic field: it is an interesting exercise because, as
we shall see, the consistent interpretation of the effective action forces a
particular choice in the family of operators previously discussed. To require
the algebraic solution (37) also for the gauge problem gives a new
constraint, fixing the freedom on the normalization factor.
\newline
We start with the operators
\[\DI\,(s)=i\s_{a}e^{\mu}_{a}[\partial_{\mu}+\frac{1}{4}(1-s_{1})\,\Omega_{\mu}
+is_{2}\, A_{\mu}],\]
\[\tilde{\DI}\,(r)=i\tilde{\s}_{a}e^{\mu}_{a}[\partial_{\mu}
-\frac{1}{4}(1-r_{1})\,\Omega_{\mu}+ir_{2}\,A_{\mu}],\]
$A_{\mu}$ belonging to a (trivial) $U(1)$ principal bundle on the manifold:
we can consider $A_{\mu}$ real (not extended to complex values) without loss
of generality.
\newline
In absence of a curved background the correct consistent determinant is given
by:
\[r_{2}=0\,\,\,;\,\,\,s_{2}=1, \]
\[r_{2}=1\,\,\,;\,\,\,s_{2}=0, \]
depending on the chirality.
\newline
The presence of $A_{\mu}$ modifies the $SL(2,C)$ variations
(\ref{eq:lou}), (\ref{eq:reed}).
The additional contributions to $t_{1}$ and $t_{2}$ are:
\EQ
\delta t_{1}\,(\lb)\,=\,\frac{1}{4\pi}\frac{1}{2-k}
\int d^{2}x \sqrt{g}[-(r_{2}+s_{2})\frac{\epsilon
^{\mu\nu}}{\sqrt{g}}\,\partial_{\mu}A_{\nu}+i(r_{2}-s_{2})D_{\mu}A^{\mu}]
\,(r_{1}+s_{1})\,\tau,
\EN
\EQ
\delta t_{2}\,(\lb)\,=\,\frac{i}{4\pi}\frac{1}{2-k}
\int d^{2}x \sqrt{g}[-(r_{2}+s_{2})\frac{\epsilon
^{\mu\nu}}{\sqrt{g}}\,\partial_{\mu}A_{\nu}+i(r_{2}-s_{2})D_{\mu}A^{\mu}]
\,(r_{1}-s_{1})\,\tau.
\EN
The relation
\EQ
\delta t_{1}\,(\lb)\,=\pm \,i\, \delta t_{2}\,(\lb) \label{eq:oppio}
\EN
immediately requires either $r_{1}=0$ or $s_{1}=0$.
Actually we can relax  condition
(\ref{eq:oppio}): as in the pure gravity case
we try to satisfy (\ref{eq:oppio})
up to coboundary terms. In so doing we add a local polynomial, depending on a
real parameter $\alpha$, to the $\zeta$-function effective action:
\EQ
Q(\alpha)\,=\,\frac{\alpha}{2-k}\int d^{2}x \sqrt{g}[\,i \frac{\epsilon^{\mu
\nu}}{\sqrt{g}}(r_{2}+s_{2})\Omega_{\mu}A_{\nu}+(r_{2}-s_{2})\Omega_{\mu}
A^{\mu}].
\EN
Under extended Lorentz rotations we obtain:
\EQ
\delta_{1}(\lb)Q(\alpha)\,=\,\frac{\alpha}{2-k}\int d^{2}x \sqrt{g}
[- \frac{\epsilon^{\mu\nu}}{\sqrt{g}}
(r_{2}+s_{2})\partial_{\mu}A_{\nu}+(r_{2}-s_{2})iD_{\mu}A^{\mu}]\tau,
\EN
\EQ
\delta_{2}(\lb)Q(\alpha)\,=\,\frac{\alpha}{2-k}\int d^{2}x \sqrt{g}
[ \, \frac{\epsilon^{\mu\nu}}{\sqrt{g}}
(r_{2}-s_{2})\partial_{\mu}A_{\nu}+(r_{2}+s_{2})\,iD_{\mu}A^{\mu}]i\tau.
\EN
The consistent solution is now encoded in the following system of equations:
\[(r_{1}+s_{1})(r_{2}+s_{2})+\alpha (r_{2}+s_{2})\,=\,\pm
[(r_{1}-s_{1})(r_{2}+s_{2})-\alpha (r_{2}-s_{2})], \]
\EQ
(r_{1}+s_{1})(r_{2}-s_{2})+\alpha (r_{2}-s_{2})\,=\,\pm
[(r_{1}-s_{1})(r_{2}-s_{2})-\alpha (r_{2}+s_{2})],
\EN
that is equivalent to
\[ r_{2}[r_{1}+s_{1}+\alpha]\,=\,\pm r_{2}[r_{1}-s_{1}-\alpha], \]
\EQ
s_{2}[r_{1}+s_{1}+\alpha]\,=\,\pm s_{2}[r_{1}-s_{1}+\alpha].
\EN
The solution $r_{2}=0$, $s_{2}=0$ is trivial (no gauge field!).
\newline
On the other hand $r_{2}\neq 0$, $s_{2}\neq 0$ gives $\alpha=0$, falling
into the case $r_{1}=0$ or $s_{1}=0$.
\newline
We are left with the two chirally symmetric solutions:
\EQ
r_{2}=0\,;\,s_{2}\, \neq 0 \,\,\, \Rightarrow s_{1}=0
\,\,\,\,\,or\,\,\,\,\,
\alpha=-r_{1},
\label{eq:hope}
\EN
\EQ
s_{2}=0\,;\,r_{2}\, \neq 0 \,\,\, \Rightarrow r_{1}=0
\,\,\,\,\,or\,\,\,\,\,
\alpha=-s_{1},
\label{eq:hape}
\EN
that fix the weight of the spin connection (linked to $r_{1}$ and $s_{1}$).
\newline
The next step is to study the gauge algebra: in order to perform the variation
of $Q(\alpha)$ we have to understand how to realize the non-compact
transformation on $A_{\mu}$ without the extension to complex values.
\newline
Let us make an imaginary gauge transformation on $\DI$ with parameter $i\beta$:
\[ A_{\mu} \rightarrow A_{\mu}+i\partial_{\mu} \beta .\]
Using the properties of the Weyl algebra representation we get:
\[ \DI \rightarrow \DI^{'}=i\s_{a}e^{\mu}_{a}[\partial_{\mu}+\frac{1}{4}
\Omega_{\mu}+i\,A^{'}_{\mu}], \]
\EQ
A_{\mu} \rightarrow A^{'}_{\mu}=A_{\mu}+ \frac{\epsilon^{\mu\nu}}{\sqrt{g}}
\partial_{\nu}\beta.
\EN
In this way we can understand the imaginary non-compact transformation of the
extended gauge group as a real transformation: this property can be easily
generalized to the non-abelian gauge theories in $d=2$.
\newline
The gauge variations of the effective action, modified by $Q(\alpha)$, are:
\[
\delta_{1}(\beta)\tilde{\Gamma}=\frac{i}{2-k}\frac{1}{4\pi}
\int d^{2}x \sqrt{g} [H_{1}-\tilde{H}_{1}](r_{2}-s_{2})\, \beta+\]
\EQ
+\frac{i}{2-k}\frac{1}{4\pi}
\int d^{2}x \sqrt{g}\,
\al [(r_{2}+s_{2})\frac{\epsilon^{\mu\nu}}{\sqrt{g}}\partial_{\mu}\Omega_{\nu}
+ i\,(r_{2}-s_{2})D_{\mu}\Omega^{\mu}]\,\beta,
\EN
\[
\delta_{2}(\beta)\tilde{\Gamma}=-\frac{1}{2-k}\frac{1}{4\pi}
\int d^{2}x \sqrt{g} [H_{1}-\tilde{H}_{1}](r_{2}+s_{2})\, \beta+\]
\EQ
+\frac{i}{2-k}\frac{1}{4\pi}
\int d^{2}x \sqrt{g}\,
\al [(r_{2}-s_{2})\frac{\epsilon^{\mu\nu}}{\sqrt{g}}\partial_{\mu}\Omega_{\nu}
+ i\,(r_{2}+s_{2})D_{\mu}\Omega^{\mu}]\,\beta.
\EN
The explicit expression is recovered after the computation of the relevant
Seeley-de Witt coefficients (see appendix A):
\[ \delta_{1}(\beta)\tilde{\Gamma}=\frac{i}{2-k}\frac{1}{4\pi}
\int d^{2}x \sqrt{g}\, \{[-(r_{2}+s_{2})\frac{\epsilon^{\mu\nu}}{\sqrt{g}}
\partial_{\mu}A_{\nu}+ i\,(r_{2}-s_{2})D_{\mu}A^{\mu}]
\,(r_{2}-s_{2})\,\beta\,+\]
\[+\al[(r_{2}+s_{2})\frac{\epsilon^{\mu\nu}}{\sqrt{g}}\partial_{\mu}A_{\nu}
+ i\,(r_{2}-s_{2})D_{\mu}A^{\mu}]\,\beta+ \]
\EQ
+[\frac{1}{4}\,(r_{1}-s_{1})\,R-(r_{1}+s_{1})D_{\mu}\Omega^{\mu}]\,(r_{2}-s_{2})
\,\beta \}\, ,
\label{eq:feni}
\EN
\[ \delta_{2}(\beta)\tilde{\Gamma}=-\frac{1}{2-k}\frac{1}{4\pi}
\int d^{2}x \sqrt{g}\, \{[(r_{2}+s_{2})\frac{\epsilon^{\mu\nu}}{\sqrt{g}}
\partial_{\mu}A_{\nu}+ i\,(r_{2}-s_{2})D_{\mu}A^{\mu}]
\,(r_{2}+s_{2})\,\beta\,+\]
\[+\al[(r_{2}-s_{2})\frac{\epsilon^{\mu\nu}}{\sqrt{g}}\partial_{\mu}A_{\nu}
+ i\,(r_{2}+s_{2})D_{\mu}A^{\mu}]\,\beta+ \]
\EQ
+[\frac{1}{4}\,(r_{1}-s_{1})\,R-(r_{1}+s_{1})D_{\mu}\Omega^{\mu}]\,(r_{2}+s_{2})
\,\beta\}\, .
\label{eq:sido}
\EN
We are able now to solve the consistent constraint:
\EQ
\delta_{1}(\beta)\tilde{\Gamma}=\pm\,i\,\delta_{2}(\beta)\tilde{\Gamma}.
\label{eq:tiro}
\EN
The value of $r_{2}$ ($s_{2}$) can be computed directly from the
$\al$-indipendent
part: to satisfy (\ref{eq:tiro}) on the $A_{\mu}$ dependent part
of (\ref{eq:sido}), (\ref{eq:feni}) implies
\[ (r_{2}+s_{2})(r_{2}-s_{2})=\mp(r_{2}+s_{2})^{2} ,\]
\[ (r_{2}+s_{2})(r_{2}-s_{2})=\mp(r_{2}-s_{2})^{2} .\]
The only non trivial solutions, consistent with (\ref{eq:hope}) and
(\ref{eq:hape}), are:
\EQ
r_{2}=0,
\EN
\EQ
s_{2}=0.
\EN
Collecting the informations of the two different cohomological problems
(gauge and gravitational) we obtain:
\EQ
r_{2}=0\,\,\, ; \,\,\, s_{1}=0,
\EN
\EQ
s_{2}=0\,\,\, ; \,\,\, r_{1}=0.
\EN
We have made this choice to exploit the freedom on $\al$: we take respectively
\[ \al_{1}=-\frac{r_{1}}{4\pi},\]
\EQ
\al_{2}=-\frac{s_{1}}{4\pi}.
\EN
In this way one can check that the Lorentz anomaly depends only on the
spin-connection and the gauge anomaly only on the $U(1)$ connection:
\EQ
t_{1}(\lb)=\frac{i}{192\pi}\int d^{2}x \sqrt{g}\,R\, \tau+\frac{1}{4\pi}
\frac{(r_{1}-s_{1})^{2}}{2-k}\int d^{2}x \sqrt{g}\,D_{\mu}\Omega^{\mu}\,
\tau,
\EN
\EQ
a_{1}(\beta)=\frac{i}{4\pi}\frac{(r_{2}-s_{2})}{2-k}\int d^{2}x
\sqrt{g}\,
[-(r_{2}+s_{2})\frac{\epsilon^{\mu\nu}}{\sqrt{g}}\partial_{\mu}A_{\nu}+i
(r_{2}-s_{2})D_{\mu}A^{\mu}]\,\beta.
\EN
The choice $s_{1}=0$ ($r_{1}=0$) forces on (\ref{eq:zaza})
(\ref{eq:zuzu})) $k=1$: the last unknown parameter
is $s_{2}$ ($r_{2}$). They are fixed by the normalization condition for the
$U(1)$ dependent part of the effective action: it is an exercise, using the
decoupling technique discussed in appendix C, to compute the determinant for
generic $r_{2}$ and $s_{2}$ and, comparing the result with the one for
its modulus, to find:
\EQ
s_{2}=1\,\,\,(r_{2}=1)
\EN
The correct determinant is (we are using the second type of solution):
\[ \tilde{\Gamma}[\Omega,A]= \ln Det[\tilde{\DI}\,
(\frac{\sqrt{3}}{3};0)\DI\,(0;1)]\]
\[=\frac{1}{192\pi}\int d^{2}x\sqrt{g(x)}\int d^{2}y\sqrt{g(y)}
[R(x)\Delta^{-1}_{g}(x,y)R(y)+
iR(x)\Delta^{-1}_{g}(x,y)\frac{1}{\sqrt{g}}\partial_{\nu}(\sqrt{g(y)}
\Omega^{\nu}(y))] \]
\[-\frac{1}{4\pi}\int d^{2}x\sqrt{g(x)}\int d^{2}y\sqrt{g(y)}
\frac{\epsilon^{\mu\nu}}{\sqrt{g(x)}}\partial_{\mu}A_{\nu}\Delta^{-1}_{g}(x,y)
(\frac{\epsilon^{\rho\sigma}}{\sqrt{g(y)}}+ig^{\rho\sigma})\partial_{\rho}
\sqrt{g(y)} A_{\sigma}\]
\[-\frac{1}{24\pi}\int d^{2}x\sqrt{g}\,\Omega_{\mu}\Omega^{\mu}
-\frac{1}{8\pi}\int d^{2}x\sqrt{g}\,A_{\mu}A^{\mu}\]
\EQ
+\frac{1}{4\pi}
\int d^{2}x\sqrt{g}\,[i\frac{\epsilon^{\mu\nu}}{\sqrt{g}}\Omega_{\mu}
A_{\nu}-i\,D_{\mu}A^{\mu}].
\EN
A suitable introduction of local polynomials brings this expression to the
minimal form given by Leuterwyler in \cite{rfi}. We notice that the operator
is completely fixed and the flat case is smoothly connected to the curved one:
in other words the operator realizing the Weyl effective action on the curved
space reduces to the correct one (for gauge theories) in the limit $e^{\mu}_{a}
\rightarrow\, \delta^{\mu}_{a}$. The symmetry between the inequivalent
representation of the Weyl algebra is expressed changing $r_{i}$ in $s_{i}$.
\vfill\eject

\section{Conclusions}
In the case of gauge theories we have shown how to describe consistent and
covariant anomalies in a unified scheme: the problem of representing
this solutions in a functional approach has been completely understood,
and it is not difficult to extend the results to non trivial fiber
bundles. The essential role of the complexification of the gauge group
appears in both algebraic and functional framework.
The gravitational case is more involved,
for the presence of two symmetries and a more sophisticated geometrical
background. Simplifying the problem to the pure Lorentz anomaly
we have studied the covariant sector and we have shown how it is related
to the complex $SO(2n,C)$ extension of $SO(2n,R)$. Unfortunately we have
not been able to find a functional representation for the consistent
vacuum functional in $d=2n$. Nevertheless the simplicity of $d=2$ has
allowed
us to calculate exactly the determinant, showing that a  continuous family of
operators, admitting as determinant the Weyl one, does exist. In some sense
they correspond to different regularizations of the theory, changing
only for local terms in their effective action. Our final results are in
agreement with the Leutwyler's calculations \cite{rfi}, which are
nevertheless based on a different philosophy and procedure.
\vskip 1.0 truecm
I would like to thank Prof. Antonio Bassetto for reading the manuscript and
for many useful comments. I am also grateful to Prof. Roberto Soldati and
Dott. Paola Giacconi for several critical and stimulating discussions on
the subject.

\vfill\eject

\appendix
\section{Heat-kernel expansion}
Let us consider a second order differential operator
\EQ
A=-g^{\mu\nu}\partial_{\mu}\partial_{\nu}+f^{\mu}\partial_{\mu}+h.
\EN
The fields $f$ and $h$ are allowed to be matrix-valued: we assume that
$g^{\mu\nu}$ is positive and $A$ acts on a compact manifold: $A$ can be
rewritten in the form
\EQ
A=-\frac{1}{\sqrt{g}}(\partial_{\mu}+\tilde{v}_{\mu})\sqrt{g}\,
g^{\mu\nu}(\partial_{\nu}+v_{\nu})+V.
\EN
The heat-kernel of $A$ is a solution of:
\EQ
\frac{\partial}{\partial t}K(x,y;t)+A_{x}K(x,y;t)=0,
\EN
with the boundary condition:
\EQ
K(x,y;0)=\de^{d}(x-y)\frac{1}{\sqrt{g}}.
\EN
We remark that $K(x,y;t)$ also exists for operators which are
not positive definite
\cite{rfs}: the important point is the positivity of the metric
itself, that controls the asymptotic behaviour of the eigenvalues of
$A$. We denote the length of the shortest path from $x$ to $y$ by
$\sqrt{\eta(x,y)}$ (geodesic distance): for $t\rightarrow 0$ $K$ admits
the expansion  \cite{rfh}:
\EQ
K(x,y;t)\rightarrow(4\pi)^{-\frac{d}{2}}\exp [-\frac{\eta(x,y)}{4t}]
\sum_{n=0}^{\infty}t^{n}H_{n}(x,y).
\EN
Using the property:
\EQ
g^{\mu\nu}\partial_{\mu}\eta\partial_{\nu}\eta=4\eta
\EN
one can verify that the heat-kernel equation is satisfied order by order
in $t$, provided that the coefficients obey to the recursive differential
equation:
\EQ
\frac{1}{2}\partial^{\mu}\eta D_{\mu}H_{n}+(n+\frac{1}{4}D^{\mu}D_{\mu}\eta-
\frac{d}{2})H_{n}=-AH_{n-1}  \label{eq:totti}
\EN
where $D_{\mu}$ is built with the field $v_{\mu}$ and the Levi-Civita
connection $\Gamma^{\lb}_{\mu\nu}$.
\newline
This differential equation implies that, at $x=y$, $H_{n}$ and their
derivatives reduce to local polynomials. Using the property:
\EQ
(D_{\mu}D_{\nu}-D_{\nu}D_{\mu})C_{\al}=R^{\lb}_{\al\mu\nu}C_{\lb}
+f_{\mu\nu}C_{\al}
\EN
it is possible to express all the quantities in terms of the curvatures
$R^{\al}_{\beta\mu\nu}$, $f_{\mu\nu}=\partial_{\mu}v_{\nu}-\partial_{\nu}
v_{\mu}+[v_{\mu},v_{\nu}]$ and the derivatives of $\eta$. It is a matter
of labour to obtain the heat-kernel coefficients $H_{n}$: as an
example we report the computation of $H_{1}$ for the operator
\[A=\tilde{\DI}\,(r)\DI\,(s)=\tilde{\s}_{a}e^{\mu}_{a}[i\partial_{\mu}
+(1-r)\Omega_{\mu}]\,
\s_{a}e^{\nu}_{a}[i\partial_{\nu}-(1-s)\Omega_{\nu}]\]
that is the relevant one in the case of $d=2$ gravitational determinant
($\Omega_{\mu}=\frac{1}{4i}\Omega_{\mu ab}\tilde{\s}_{a}\s_{b}$). Using
the recursive differential equation at $x=y$ it is easy to prove:
\EQ
H_{1}=\frac{1}{6}R-V.
\EN
Putting $\tilde{\DI}\,(r)\DI\,(s)$ in the ``canonical'' form:
\[v_{\mu}=\frac{1}{2}[(1-r)\Omega_{\mu}+(1-s)\Omega_{\mu}+2\Omega_{\mu}]+\,
\frac{1}{2}\,\frac{i}{\sqrt{g}}\epsilon^{\lb\nu}g_{\mu\nu}[(r-1)\Omega_{\lb}+
2\Omega_{\lb}+(s-1)\Omega_{\lb}],\]
\EQ
\tilde{v}_{\mu}=-v_{\mu},
\EN
and
\EQ
V=\frac{1}{8}(2-s+r)R+\frac{i}{2}(r+s)D_{\mu}\Omega^{\mu}.
\EN
Hence
\EQ
H_{1}(\tilde{\DI}\,(r)\DI\,(s))=\frac{1}{6}R-\frac{1}{8}(2+r-s)R
+\frac{i}{2}(r+s)D_{\mu}\Omega^{\mu}.
\EN
To get $\tilde{H}_{1}$ we have to send $r\rightarrow s$ and
$\Omega_{\mu}\rightarrow -\Omega_{\mu}$
All the heat-kernel coefficients used in the various chapters were
obtained in an analogous way.
\section{The variation of the determinant}
In term of the heat-kernel $K(x,y;t)$ we can express the $\zeta$-
function connected to the operator $A$:
\EQ
\zeta(s;A)=Tr[A^{-s}]=\frac{1}{\Gamma(s)}\int_{M}dx\int_{0}^{\infty}\,dt
\,t^{s-1}tr[K(x,x;t)]
\EN
taking the relation between kernel:
\EQ
<x|A^{-s}|y>=\frac{1}{\Gamma(s)}\int_{0}^{\infty}\,dt
\,t^{s-1}tr[K(x,y;t)] \label{eq:titti}
\EN
into account.
\newline
If we perturb the operator $A$ with $\epsilon A_{1}$, $\epsilon\rightarrow
0$, where $A_{1}$ is a differential operator
($ordA\geq ordA_{1}$), we obtain [6]:
\[ \zeta(s;A+\epsilon A_{1})=\zeta(s)+\epsilon f(s)+O(\epsilon^{2}),\]
\EQ
f(s)=sTr[A^{-s-1}A_{1}].
\EN
{}From the definition of determinant [6]:
\EQ
-\ln DetA=\frac{d}{ds}\zeta(s)|_{s=0}
\EN
one recovers:
\EQ
-\ln Det(A+\epsilon A_{1})=-\ln
DetA+\epsilon\frac{d}{ds}(sTr[A^{-s-1}A_{1}])_{s=0}.
\EN
Using the relation (\ref{eq:titti}) and the expansion (\ref{eq:totti}),
together with the fact we are working with trace-class operators, we
could in principle compute all variations. The fact that in
most of our computations only one heat-kernel coefficient is needed,
is due to the particular form of the variations:
\[A_{1}=\al(x)A+A\beta(x).\]
In this case it is rather clear that only one coefficient contributes
in the limit $s=0$.
\section{Decoupling technique}
We want to show how the decoupling technique works in the calculation of
(\ref{eq:fre}). The relevant finite transformation is:
\bea
G\rightarrow G(1-y)=G_{y}\nonumber\\
F\rightarrow F(1-y)=F_{y}
\eea
Let us define the quantity:
\EQ
\Gamma'[y;r,s]=\lim_{\de y\rightarrow 0}\frac{1}{\de y}(\Gamma[y-\de
y;r,s]-\Gamma[y;r,s])
\EN
where
\EQ
\Gamma[\xi;r,s]=-\frac{1}{2}\frac{d}{dt}
Tr[(\tilde{\DI}\,(r;\xi)\DI\,(s;\xi))^{-t}]_{t=0}.
\EN
$\tilde{\DI}\,(r;\xi)$ and $\DI\,(s;\xi)$ are obtained from the usual
$\tilde{\DI}\,(r)$ and $\DI\,(s)$ by the substitution
\[G\rightarrow (1-\xi)G,\]
\[F\rightarrow (1-\xi)F.\]
It is clear that
\bea
\Gamma[1;r,s]=\Gamma[r,s],\nonumber\\
\Gamma[0;r,s]=-\ln Det\partial^{2},
\eea
giving:
\EQ
\int_{0}^{1}dt \Gamma'[y;r,s]=\Gamma[r,s]-\ln Det\partial^{2}.
\EN
Using the expression of $\tilde{\DI}\,(r)$ and $\DI\,(s)$ we obtain in
conformal coordinates
\[\Gamma'[y;r,s]=\frac{1}{2}Tr\{(\tilde{\DI}\,(r;y)\DI\,(s;y))^{-t}
[(2+r-s)G-i(r+s)F]+\]
\EQ
+(\DI\,(s;y)\tilde{\DI}\,(r;y))^{-t}[(2-r+s)G+i(r+s)F]\}_{t=0}.
\EN
{}From the heat-kernel expansion we get:
\[\Gamma'[y;r,s]=\frac{1}{8\pi}\int
\sqrt{g(y)}(H_{1}(\tilde{\DI}\,(r;y)\DI\,(s;y))[(2+r-s)G-i(r+s)F]+\]
\EQ
+(\tilde{H}_{1}(\DI\,(s;y)\tilde{\DI}\,(r;y))[(2-r+s)G+i(r+s)F])
\EN
with $\sqrt{g(y)}=\exp(4G(1-y))$.
\newline
The explicit calculation of $H_{1}$ and $\tilde{H}_{1}$ in isothermal
coordinates gives
\[H_{1}=\exp(-4(1-y)G)[(-4(1-y)\partial_{\mu}\partial_{\mu}G)[-\frac{1}{12}-
\frac{1}{8}(r-s)]-\frac{i}{2}(1-y)\partial_{\mu}\partial_{\mu}F(r+s)],\]
\[\tilde{H}_{1}=\exp(-4(1-y)G)[(-4(1-y)\partial_{\mu}\partial_{\mu}G)
[-\frac{1}{12}-\frac{1}{8}(s-r)]+
\frac{i}{2}(1-y)\partial_{\mu}\partial_{\mu}F(r+s)].\]
A straightforward algebra gives:
\[\Gamma'[y;r,s]=(1-y)\frac{1}{8\pi}\int d^{2}x[\frac{1}{3}(4G\partial_{\mu}
\partial_{\mu}G)-\frac{1}{4}(r-s)^{2}(4G\partial_{\mu}\partial_{\mu}G)-\]
\EQ
-\frac{i}{2}(r^{2}-s^{2})(4F\partial_{\mu}\partial_{\mu}G)+(2+r-s)^{2}
(F\partial_{\mu}\partial_{\mu}F)]
\EN
and
\[\int_{0}^{1}dy\Gamma'[y;r,s]=\frac{1}{16\pi}\int\,d^{2}x(4G\partial_{\mu}
\partial_{\mu}G)(\frac{1}{3}-\frac{1}{4}(r-s)^{2})-\]
\EQ
-\frac{i}{2}(r^{2}-s^{2})
(4F\partial_{\mu}\partial_{\mu}G)+(2+r-s)^{2}(4F\partial_{\mu}\partial_{\mu}F).
\EN
\vfill\eject

\end{document}